\documentclass[a4paper,11pt]{article}
\pdfoutput=1 % if your are submitting a pdflatex (i.e. if you have
\usepackage[skip=10pt plus1pt, indent=5pt]{parskip}
\usepackage{jcappub} % for details on the use of the package, please
\usepackage[T1]{fontenc}

\usepackage{parskip}
\usepackage{mathtools}
\usepackage{enumitem}
\usepackage{dsfont}
\usepackage{xcolor}
\definecolor{greyforboxes}{RGB}{235,235,235}

\newcommand{\Lag}{\mathcal{L}}
\newcommand{\mS}{\mathcal{S}}
\newcommand{\mT}{\mathcal{T}}
\newcommand{\upsilonh}{\hat{\Upsilon}}
\newcommand{\GB}{\mathcal{G}}

\newcommand{\nablab}{\mathring{\nabla}}
\newcommand{\Rb}{\mathring{R}}
\newcommand{\Gb}{\mathring{G}}
\newcommand{\cE}{\ensuremath{\mathcal{E}}}
\newcommand{\cL}{\ensuremath{\mathcal{L}}}
\newcommand{\Ekg}{\ensuremath{\mathcal{E_\textrm{KG}}}}

\newcommand\eN{\mathrm{e}}
\newcommand\iN{\mathrm{i}}
\newcommand\dd{\mathrm{d}}
\newcommand{\mpl}{M_{\rm Pl}}

\newcommand\lrsq[1]{\left[#1\right]}
\newcommand\lr[1]{\left(#1\right)}

\usepackage{tikz}                            %Antes de 10,10,2016 tenia aqui {pstricks}
\usepackage[framemethod=tikz]{mdframed}      % linewidth=0.4 (default)
\mdfsetup{splittopskip=\topskip, linewidth=0.5}
            
\newenvironment{boxmathdark}[1]
    {\begin{mdframed}[topline=false,rightline=false, bottomline=false,
                      linewidth=2mm, linecolor=black, backgroundcolor=greyforboxes, 
                      innertopmargin = 8pt, innerbottommargin = 8pt,
                      skipabove = 8pt, skipbelow = 8pt]}
    {\end{mdframed}}
\newenvironment{boxmathclear}[1]
    {\begin{mdframed}[topline=false,rightline=false, bottomline=false,
                      linewidth=2mm, linecolor=greyforboxes, 
                      innertopmargin = 8pt, innerbottommargin = 8pt,
                      skipabove = 8pt, skipbelow = 8pt]}
    {\end{mdframed} }

\newcommand{\lecture}[1]{%
  \refstepcounter{section}%
  \section*{Lecture \thesection. #1}}

\title{Instabilities in field theories\\
\Large Lecture notes with a view into modified gravity}

\author{Adri\`a Delhom,}
\author{Alejandro Jim\'enez-Cano}
\author{and Francisco Jos\'{e} Maldonado Torralba}

\affiliation{Laboratory of Theoretical Physics, Institute of Physics, University of Tartu, W. Ostwaldi 1, 50411 Tartu, Estonia}

\emailAdd{adria.delhom@gmail.com}
\emailAdd{alejandro.jimenez.cano@ut.ee}
\emailAdd{fmaldo01@ucm.es}

\abstract{
    Modified theories of gravity usually present new degrees of freedom, as well as higher order derivatives, wrong signs in certain terms and complicated couplings already present in the Lagrangian from the beginning or originated by the field redefinitions needed to reach an Einstein frame. As a consequence, they are very prone to present dynamical instabilities that could spoil any attempt to construct viable models within these frameworks. In these three lectures we introduce the most common types of instabilities that appear in field theory as well as some techniques to detect them, and supplement these contents with several examples. The goal is to understand the implications of having such pathological behaviors and the application of these notions to modified theories of gravity.\\
    
    \noindent{\footnotesize This text is an extended and polished version of the lectures prepared for the course `Selected Topics in the Theories of Gravity', given at the Institute of Physics (University of Tartu, Estonia) in spring 2022. Recordings of such course can be found in \url{https://button.ut.ee/b/lau-a4e-8eb-es1}.}
    
    \noindent{\footnotesize The last lecture contains part of the results in \cite{BeltranJimenez:2019hrm, BeltranJCA2021, BeltranDelhom2019, BeltranDelhom2020}}.
    
}

\begin{document}
\maketitle

\newpage

\lecture{Introduction}

One of the most relevant contributions from Newton was to provide a mathematical framework in which we can formulate problems which involve time evolution of some physical quantity. Time evolution should be intuitively understood as the process in which the values of certain variables change, as time goes from one instant to the next one.  Modern attempts to give a fundamental description of our universe are formulated as what is known as \textit{field theories}. These are simply theories where the physical information is encoded in functions, called \textit{fields}, over some base space. Depending on the formulation of the theory, the base space might be the spacetime, the phase space, or other sets with the required structure. This framework has been useful as well beyond the regime of fundamental physics, being able to provide accurate descriptions of more complex systems, finding important applications in more applied fields such as fluid dynamics, solid state physics, or any kind of system in which physical information has to be encoded in terms of a continuous variable.

From the mathematical perspective, in order to formulate a field theory, first one needs to provide a base space, in which the fields take values, and one usually expects it to have some nice mathematical properties that allow to introduce \textit{derivatives},\footnote{
   These properties are related to the topological and smooth structures of the base space, typically being a smooth manifold.}
the basic mathematical objects needed to define evolution in a generalized sense. In this framework, now one can think of `evolution' in space as well as in time, as one can think of the ways in which a mathematical function over spacetime changes from point to point both in spatial and time directions. In fact, according to Relativity, there are no such things as absolute spatial and time directions. Hence, in order to describe time evolution in a field theory, one must first decide what direction in spacetime will be understood as time direction. This generally provides a splitting of spacetime into space + time in terms of a succession of hypersurfaces that cover the whole spacetime and do not intersect.\footnote{If you want to understand this more precisely, look up `global hyperbolicity' and `3+1 decomposition'.} These hypersurfaces play the role of instants, and they can be used to define time evolution as the change undergone in physical quantities when passing from one of this surfaces to the next one. In general, when we use the word \textit{evolution}, we will be referring to time evolution in this sense.

Having this base space, and the physically relevant functions over it to be described, the last ingredient to define the theory is something that establishes how the physical quantities evolve once we know their value at an instant of time (this is usually called \textit{initial value problem}) or if we know their value at a suitable region of space (usually called \textit{boundary value problem}). Partial differential equations (PDEs) are exactly the mathematical objects that we are looking for: they can provide a value for physical fields over the whole spacetime provided that we specify initial data or suitable boundary conditions. Thus, from a mathematical perspective, when dealing with field theories we are just dealing with a system of PDEs called \textit{field equations} that describe a set of (physically relevant) functions over spacetime (or some other related space of interest). When one specifies the values at each point for one of these fields, we call that a particular \textit{configuration} of the field. In modern theories, the field equations are usually obtained by assuming that the physical field configurations (i.e., the solutions to the field equations) are those which correspond to critical points of a certain functional of the fields that we call the \emph{action}.

Once the field theory is specified, namely, once one has defined the set of fields to be described, the space they live on, and the PDEs that govern their behavior, the fundamental question to be answered is: what are the physically relevant configurations for these fields, and what are their properties? A general answer would be that any solution to the field equations is a candidate to be a physically viable configuration according to the theory. However, we can go further, and classify the properties of these solutions, regarding the type of initial/boundary data that gives rise to them, their asymptotic properties, and/or their compatibility with observed properties of the universe. Indeed, from a physical perspective, we know that we can never determine initial conditions exactly, so we might be wondering what happens if instead of having some initial data that determines an exact solution, we have initial data close to it. This can also be formulated by the question: what happens if we perform small perturbations of a given exact solution? Do the deviations with respect to the unperturbed solution remain arbitrarily small for sufficiently small perturbations, or does the perturbed evolution lead to a completely different physical system? In short, in the first case we say that such solution is stable under small perturbations, and in the second case we say that such solution is unstable under small perturbations (or perturbatively stable/unstable respectively). Thus, perturbative stability will be characterized by assessing whether perturbations remain small through the whole evolution, or their corrections become comparable to the `size' of the exact solution, thus pointing that these perturbations destabilize it\footnote{This can be extended to any observables related to that solution: If a small perturbation changes the value of an observable substantially after time evolution, then one must say that such observable is not stable under perturbations.}. This view gives a broad but imprecise idea of what instabilities are. The goal of these lectures is to provide a more precise description of these issues and, after that, to discuss their relevance in some modified gravity scenarios.

\addcontentsline{toc}{section}{Introduction}

\subsection{Warm up examples}

In this section we aim to give a more precise definition of the basic types of instabilities arising in field theory but, first, let us provide with a little more context to the reader. In physics we try to provide a mathematical description to observed physical systems. One key characteristic of these systems is to understand whether producing small modifications on them leads to small changes or completely destroys the system leading to other kind of phenomenology. For instance, imagine that one has a bound state of a planet orbiting a star, described by an exact solution to the two body problem. Now, one can decide to perturb this planet by,  e.g., hitting it with a small asteroid. Will the orbit undergo a small change or, on the contrary, the planetary orbit will be destabilized sending it to infinity (or inside the star) and completely destroy the 2-particle bound state previously formed by planet-star? This corresponds to a physical example of the question {\it do the deviations with respect to the unperturbed solution remain arbitrarily small for sufficiently small perturbations, or does the perturbed evolution lead to a completely different physical system?} that we described, in more mathematical terms, in the previous section. Now, note that, if a system is strictly unstable, the probability that it forms and stays that way given some arbitrarily small set of initial conditions tends to zero. Therefore, one property that we require for the physical systems that we observe is that they are stable under perturbations which are small enough. Another question that might arise in the case of unstable systems is: how long does it take for the system to break apart? If the instability time is long enough, this system may be physically allowed only in some time window smaller than the instability time.

To be clear: the question of stability refers to exact solutions of a theory, often called \textit{vacua} or \textit{backgrounds}, and it has to be studied in general case by case. Sometimes, when there are generic instabilities that arise for any vacuum of the theory, or when a background that is of physical relevance (e.g. Minkowski, de Sitter, etc.) is unstable, we say that the theory is unstable. But, strictly, stability is a property of exact solutions, not theories. Having said this, let us present some simple examples that will allow us to better understand how the analysis of these issues works, and to provide a classification of common types of instabilities arising in field theory.

\begin{boxmathclear}

\noindent{\bf Example: Free scalar field around Minkowski.}

A free linear relativistic scalar field $\phi(t,\vec{x})$ is described by the real Klein-Gordon (KG) Lagrangian
\begin{equation}
    \mathcal{L}_{\textrm{KG}}=\frac{1}{2}\partial^\mu\phi\partial_\mu \phi-\frac{1}{2}m^2\phi^2
\end{equation}
which, upon extremizing the associated action functional, leads to the following field equations
\begin{equation}
    \Ekg(\phi)\equiv \ddot{\phi}-\Delta\phi+m^2\phi=0
\end{equation}
 Here the dots represent (partial) time derivatives, $\Delta$ is the standard Laplacian in $\mathbb{R}^3$ and $m^2$ is a constant. The general solution of this PDE can be expressed as follows in terms of Fourier modes,
\begin{equation}
    \phi_0(t,\vec{x})= \int \dd^3 k \Big[ A_k\, \eN^{-\iN(\vec{k} \cdot \vec{x}-\omega t)} + A_k^\ast\, \eN^{\iN(\vec{k}\cdot\vec{x}-\omega t)}\Big]_{\omega=\omega(\vec{k})},
\end{equation}
where
\begin{equation}
    \omega(\vec{k}):=+\sqrt{|\vec{k}|^2+m^2},
\end{equation}
and for arbitrary constants $A_k$. Provided that $m^2$ is real and non-negative, the integrand is an oscillatory (or constant) function which remains bounded over spacetime, namely for each mode there is a real constant $C_k$ such that $|\phi(t,\vec{x})|<C_k\ \forall (t,\vec{x})$ (actually any $C_k>\sqrt{2}|A_k|$). 

The stress energy tensor of the real free scalar field can be computed from the above Lagrangian. From there, one can obtain the energy density of a given field configuration for a given observer as the 00 component of the stress energy tensor in coordinates $(t,\vec{x})$ adapted to such observer yielding
\begin{equation}
    T_{00}=\frac{1}{2}\lrsq{\dot{\phi}^2+(\vec{\nabla}\phi)^2+m^2\phi^2}
\end{equation}
We see that this quantity is also non-negative provided that $m^2\geq 0$. Note that this condition is crucial for having both oscillatory and positive-definite energy-density solutions to the field equations. Indeed, if $m^2<0$, then we might have exponentially growing/decaying solutions. While exponential decay does not destabilize a background, exponential growth is a typical cause of background instabilities due to unbounded growth of perturbations, as we will see later.

Having found exact solutions $\Ekg(\phi_0)=0$, we can assess the stability of these solutions by studying how small perturbations of such backgrounds evolve. Consider an arbitrary (small) perturbation of such a solution, $\phi(t,\vec{x})=\phi_0(t,\vec{x})+\varphi(t,\vec{x})$; in this case, since the equation (in other words, the differential operator $\Ekg$) is linear, $\Ekg(\phi)=\Ekg(\phi_0)+\Ekg(\varphi)=\Ekg(\varphi)$, so that from $\Ekg(\phi)=0$ we obtain a field equation for the perturbations which is exact to all orders and is just the original KG equation, so that we already know the form of the solutions for the perturbations

\begin{equation}
\varphi_0(t,\vec{x})= \int \dd^3q \Big[ a_q\eN^{-\iN(\vec{k}\cdot\vec{q}-\omega t)}+a^\ast_q\eN^{\iN(\vec{q}\cdot\vec{x}-\omega t)}\Big]_{\omega=\omega(\vec{q})}   
\end{equation}
Assuming that we perturb weakly each mode, namely that $|a_k/A_k|\ll 1$ at initial times, we have that $|\varphi_0/\phi_0|\ll 1$ will be true at all times, guaranteeing the stability of the particular exact solutions $\phi_k$ as well as the general one, and therefore the validity of the perturbative expansion at all times. This is a trivial example of how to check whether an solution is stable: the key point is that the perturbative expansion is well defined at all times so that the deviations from the exact solution are controlled by some small parameter prevents perturbations to grow to the `size' of the original solution itself. 

\end{boxmathclear}

In the above example, the field equations for the perturbations can actually be derived to all orders due to linearity (wave superposition). This is a very particular feature of this example and will not arise in general cases. Indeed, in order to allow for interesting background solutions, one usually requires a system with various fields and/or nonlinearities to be present. In that case, the superposition principle will not apply, and one indeed has to compute the perturbation equations order by order. Let us warm up with a simple nonlinear example which allows to show this better:

\begin{boxmathclear}

\noindent{\bf Example: Scalar field with a cubic interaction.}

\noindent {\footnotesize Note: Throughout this example we will omit the derivatives in the functional dependence (e.g., $\mathcal{L}(\phi)\equiv\mathcal{L}(\phi, \partial_\mu \phi)$) to alleviate the notation.}

Consider the following real scalar Lagrangian
\begin{align}\label{eq:TestLag1}
    &\mathcal{L}(\phi)=\frac{1}{2}\partial^\mu\phi\partial_\mu\phi-\frac{1}{2}m^2\phi^2+\frac{\lambda}{2}\phi\partial^\mu\phi\partial_\mu\phi.
\end{align}
Its field equations read
\begin{equation}
    \cE(\phi)\equiv\Box\phi+\frac{\lambda}{2(1+\lambda\phi)}\partial^\mu\phi\partial_\mu\phi+\frac{m^2}{1+\lambda\phi}\phi=0.
\end{equation}
These equations can have several branches of exact nontrivial solutions.\footnotemark{} In particular, assuming $\partial_i\phi=0$ (namely, homogeneity and isotropy), there exist nontrivial solutions $\phi_0(t)$. Now, we want to understand how perturbations behave around this background. To that end, consider $\phi(t,\vec{x})=\phi_0(t)+\varphi(t,\vec{x})$. In full generality, we can expand the above field equations $\varphi$, obtaining in general 
\begin{equation}
\cE(\phi)=\cE(\phi_0)+\sum_{n=1}^{\infty}\cE^{(n)}_{\phi_0}(\varphi),
\end{equation}
where $\cE^{(n)}_{\phi_0}(\varphi)$ is of order $\varphi^n$ and depends on the background $\phi_0$ around which the perturbative expansion is considered. Particularly, we obtain for linear perturbations the field equation
\begin{equation}
    \cE_{\phi_0}^{(1)}(\varphi)\equiv\ddot{\varphi}-\Delta\varphi+\frac{\lambda\dot\phi_0}{1+\lambda\phi_0}\dot\varphi+\frac{2m^2-\lambda^2\dot\phi_0^2}{2(1+\lambda\phi_0)^2}\,\varphi=0.
\end{equation}

We can see that linear perturbations $\varphi(t,\vec{x})$ for \eqref{eq:TestLag1} around the vacuum $\phi_0$ behave as wave like perturbations interacting with the background through a linear term in $\dot\phi$ and an effective mass term. The former will act as a damping/amplifying term depending on the sign of the background-dependent coefficient $(1+\lambda\phi_0)^{-1}\lambda\dot\phi_0$. The sign of the effective mass term can also qualitatively change the behavior of the solutions as we will see later. 

Note that the equations for linear perturbations can also be obtained from expanding the Lagrangian in perturbations as
\begin{equation}
    \cL(\phi_0+\varphi)=\cL(\phi_0)+\sum_{n=1}^{\infty}\cL_{\phi_0}^{(n)}(\varphi),
\end{equation}
where $\cL^{(n)}_{1,\phi_0}(\varphi)$ is of order $\varphi^n$ and depends on the background $\phi_0$ around which the perturbative expansion is considered. Now, because field equations are obtained taking first derivatives of the Lagrangian with respect to the corresponding field, in order to obtain the equations for perturbations up to order $n$ we need to expand the Lagrangian up to order $(n+1)$ in perturbations. Hence, in order to obtain the equations describing linear perturbations, we need to expand $\cL(\phi_0+\varphi)$ up to quadratic order in $\varphi$, finding
\begin{align}
&\cL_{\phi_0}^{(1)}(\varphi)=\lr{1+\lambda\phi_0}\dot\phi_0\dot\varphi+\lr{\frac{\lambda}{2}\dot\phi_0^2-m^2\phi_0}\varphi,\\
&\cL_{\phi_0}^{(2)}(\varphi)=\frac{1+\lambda\phi_0}{2}\partial^\mu\varphi\partial_\mu\varphi+\lambda\dot\phi_0\dot\varphi\varphi-\frac{m^2}{2}\varphi^2.\label{eq:TestLag12pert}
\end{align}

Computing the field equations for the perturbations $\varphi$, you can check that $\cL_{\phi_0}^{(1)}(\varphi)$ yields the equations for the background $\cE(\phi_0)=0$ which are satisfied by construction, and $\cL_{\phi_0}^{(2)}(\varphi)$ yields the field equations describing linear perturbations $\cE^{(1)}_{\phi_0}(\varphi)=0$ if one writes $\ddot\phi_0$ in terms of $\phi_0$ and $\dot\phi_0$ using the background equations. After integrating by parts, the Lagrangian \eqref{eq:TestLag12pert} can also be schematically written as
\begin{equation}
\cL_{\phi_0}^{(2)}(\varphi)=\frac{1}{2}G^{\mu\nu}(\phi_0)\partial_\mu\varphi\partial_\nu\varphi-\frac{\mu}{2}\varphi^2,
\end{equation}
where $G^{\mu\nu}(\phi)=(1+\lambda\phi)\eta^{\mu\nu}$ can be understood as an effective metric on top of which linear perturbations propagate and $\mu=\lambda\ddot\phi_0+m^2$ can be understood as the squared effective mass of the perturbations. You might find other example where $G^{\mu\nu}$ is not proportional to $\eta^{\mu\nu}$. Sometimes you might read the name \textit{conformal/disformal couplings} to the terms that can be encoded in an effective metric and are proportional/non-proportional to the Minkowski metric. Indeed, the same applies when the background metric is non-Minkowskian and conformal/disformal terms appear in the equation of some field.

\end{boxmathclear}
\footnotetext{You can check this using, e.g., Mathematica. For a reasonable computation time, try assuming $\partial_i\phi=0$ (namely homogeneity and isotropy) and give some particular values to the constants. You will see that a complicated nontrivial solution exists.}

We can also consider examples where the background on top of which perturbations propagate is not a background of the field perturbed, but of other field in the theory:

\begin{boxmathclear}

\noindent{\bf Example. Massless scalar field in an homogeneous and isotropic universe.}

We consider an homogeneous and isotropic universe described by the Friedman-Lema\^itre-Robertson-Walker metric, which in suitable coordinates reads
\begin{equation}
    \boldsymbol{g}_\text{FLRW}=g_{\mu\nu}\dd x^\mu \dd x^\nu=\dd t^2-a^2(t)\delta_{ij}\dd x^i \dd x^j\,,
\end{equation}
and where $a(t)$ is the scale factor which describes the expansion of the universe. The Lagrangian describing a minimally coupled massless scalar field on this cosmological background is
\begin{equation}
\cL=\frac{1}{2}\dot\phi^2-\frac{1}{2a^{2}}(\vec{\nabla}\phi)^2=\frac{1}{2}g_\text{FLRW}^{\mu\nu}\partial_\mu\phi\partial_\nu\phi
\end{equation}
which leads to the field equations
\begin{equation}
\cE_\text{FLRW}(\phi)\equiv\ddot\phi+3H\dot\phi-a^{-2}\Delta\phi=0
\end{equation}
where $H:=\dot a/a$ is the Hubble rate. Here a trivial background solution for the scalar field is a constant one, $\phi_0(t,\vec{x})=C$, and the above Lagrangian and scalar field equations are appropriate for describing linear scalar perturbations as well. Note that the sign of the Hubble rate (namely expansion/contraction, as $a(t)>0$) enters the equation as a damping/amplification term for the scalar perturbations. For the case where the perturbations get amplified because the universe is contracting, the scalar field will backreact on the metric through its coupling via the Einstein equations in a nontrivial way when the exponential growth becomes non-negligible.
\end{boxmathclear}

\subsection{Common types of instabilities and their physical implications}

The above examples are intended to give the reader a notion of what we mean by `perturbations on top of a given background (or vacuum)'. The vacua can be exact solutions of the very same field that we are perturbing or exact solutions of other fields of the theory which couple to the fields that we want to perturb. Being familiar with this notions, we see that the result is some Lagrangian or field equations which, for the scalar field, typically can be written as modifications of the KG equation where the coefficients of each term depend nontrivially on the background.

To understand the basic types of instabilities that can arise, we can study the problem from a generic perspective for a quadratic scalar Lagrangian that describes linear scalar field perturbations in an isotropic background\footnote{Although we could do the most general case without any symmetries, this case is enough to illustrate the most common types of instabilities, and isotropy facilitates the identification of the key aspects of the problem.} without amplification/damping terms.\footnote{These can also be relevant for stability, but they would obscure the analysis of the types of instabilities that we want to classify.} Our Lagrangian for scalar perturbations will thus be of the form
\begin{equation}\label{eq:EffectiveScalarLag}
\mathcal{L}=\frac{1}{2}\big(a \dot\varphi^2- b (\vec{\nabla}\varphi)^2-\mu \varphi^2\big),
\end{equation}
where the coefficients $a$, $b$, $\mu$ encode relevant information about the background. Hence, these coefficients can in general depend on spacetime coordinates, but their variations will typically occur over time/length scales $T$, $L$ that  characterize the background where the scalar field is propagating on. We want to study the behavior of perturbations on much shorter time/length scales than the ones that characterize the background. The field equations are therefore
\begin{equation}
\mathcal{E}^{(0)}(\varphi)\equiv\ddot \varphi-\frac{b}{a}\Delta\varphi+\frac{\mu}{a}\varphi=\mathcal{O}(T^{-1},L^{-1}).
\end{equation}
Here, variations of the background-dependent coefficients, which are suppressed by the scales $T$ and $L$, are sub-leading effects if the period/wavelength of the perturbations are kept small enough.\footnote{Here the superindex $(n)$ means that the exression is of order $T^{-n}$} Since, the background dependent coefficients will be constant at leading order, we can try the following ansatz for the perturbations
\begin{equation}\label{eq:SolutionKleinGordonInstabilities}
\varphi_k=A_k \eN^{-\iN\lr{\sqrt{\frac{b}{a}}\vec{k}\cdot\vec{x}-\omega t}}+A^\ast_k \eN^{\iN\lr{\sqrt{\frac{b}{a}}\vec{k}\cdot\vec{x}-\omega t}}\qquad\text{where}\qquad \omega=+\sqrt{\frac{b}{a}|\vec{k}|^2+\frac{\mu}{a}}
\end{equation}
To see how accurate it is, we can expand $\mathcal{E}^{(1)}(\varphi_k)$ in terms of $(T^{-1},L^{-1})$, and see whether the proposed ansatz solves the leading order term. Indeed we find $\mathcal{E}^{(1)}(\varphi_k)=\mathcal{O}(T^{-1},L^{-1})$, since time/space derivatives of $a,b,\mu$ are of order $T^{-1}$ and $L^{-1}$ respectively, so that the proposed ansatz is a solution to leading order. 

Now, note that the qualitative behavior of the solutions depend on the magnitude and combinations of signs of the background-dependent coefficients. While we have oscillatory solutions for $a>0, b>0, \mu\geq 0$, this is not true anymore for different sign combinations. For instance, if $\mu$ is negative and big enough, $\omega$ becomes imaginary, which will result in exponential growth of the $A^\ast_k$ term. In this case, perturbations will grow unboundedly\footnote{There is no real constant which is bigger than $|\phi_k|$ for all times in this case, no matter how small the initial amplitude of the perturbation is.} and produce backreaction effects on the background such that the variation of the background cannot be neglected anymore (namely, the perturbative expansion in negative powers of $(T,L)$ breaks down). When this happens, we say that such background is unstable for  $\varphi$-perturbations.

In order to characterize the different qualitative behavior of the instabilities that can arise in this case, let us consider systematically the different sign combinations of the background-dependent coefficients $a$, $b$, and $\mu$ and study their physical implications for $\phi$-perturbations. We will follow and extend the discussion in \cite{Rubakov2014}.

For illustrative purposes, it will suffice to consider a spatially homogeneous background slowly varying in time, though the arguments generalize in a straightforward manner. \\

\subsubsection{Stable case}

\vspace{.3cm}
\noindent{$a>0$, $b>0$ and $\mu\geq0$}

This sign configuration leads to a wave equation with similar characteristics than the KG equation: it is a hyperbolic equation with oscillatory solutions whose frequency $\omega$ is always real, and whose energy density is always positive. Given their oscillatory behavior, it is straightforward to check that  $\phi$-perturbations remain bounded at all times (and throughout space), and that they propagate at a speed given by $\sqrt{b/a}$ (in natural units). The causal character of the perturbations depends on the ratio $b/a$. If $b/a<1$ the perturbations are subluminal speeds. If $b/a>1$ the perturbations propagate at superluminal speeds.\footnote{
   Although this does not jeopardize the stability of the background, it is a sign that the corresponding field theory is not the low energy limit of a Lorentz invariant and unitary quantum theory \cite{Adams:2006sv}.}
Finally, for $b=a$ the perturbations travel at the speed of light. Although there is no problem with this, note that the smallest modification of the background (caused e.g. by the backreaction of the perturbations) could render $b>a$ and lead to superluminal propagation.\\

\subsubsection{Tachyonic instability: problems with IR modes}
\vspace{.3cm}
\noindent{$a>0$, $b>0$ and $\mu<0$}

 For this sign combination, the frequency of the modes with sufficiently small momenta becomes purely imaginary. This occurs for modes such that $|\vec{k}|<{k}_{\textrm{low}}:=\sqrt{|\mu|/b}$, as the radicand defining $\omega$ becomes negative for such modes. Therefore, though high-momentum perturbations are still oscillatory, modes with momentum lower than $k_{\textrm{low}}$ develop exponential growth or decay, since
 \begin{equation}\label{eq:Tachyonicmodes}
\varphi_k=A_k \eN^{-\iN\sqrt{\frac{b}{a}}\vec{k}\cdot\vec{x}}\eN^{-|\omega| t}+A^\ast_k \eN^{\iN\sqrt{\frac{b}{a}}\vec{k}\cdot\vec{x}}\eN^{|\omega| t} \end{equation}
where we have substituted $\omega=\iN |\omega|$. While the exponentially decaying modes are perfectly stable, the problem lies with the modes that grow exponentially, also called \textit{tachyonic} modes. Looking at \eqref{eq:Tachyonicmodes}, we see that though at early times $|\omega|t\ll 1$ the real exponentials are not relevant, at late times when $|\omega|t\gg 1$ the system is completely characterized by the exponential growth of the $\eN^{|\omega|t}$ modes, as the $\eN^{-|\omega|t}$ modes have decayed by that time. Therefore, we see that the would-be frequency of these modes does not have a physical meaning of `how many times does this system undergo a periodic motion within a given time interval', but instead, it now signals the characteristic time at which exponential growth kicks in for each of the modes, which is given by $|\omega|^{-1}$. In a system in which tachyonic modes are excited, the characteristic time at which the instability kicks in $t_c=\omega_c^{-1}$ is the one corresponding to the tachyonic mode of fastest growth. In particular, if all tachyonic modes are excited, the characteristic time of the instability is related to the effective mass of the perturbations as $\omega_c=m_{\textrm{eff}}:=\sqrt{|\mu|/a}$.  
 
Due to the presence of two scales, one characterizing the changes in the background $T$ and another characterizing the time at which the tachyonic instability becomes relevant, there could be different physical scenarios. In the case that $t_c\ll T$ the exponential growth of the tachyonic modes becomes dominant much before the background has shown any signs of evolution. Therefore, in this case, the background is unstable. Though the characteristic time of the background instability might depend on the exact way in which $\phi$ backreacts on it, it will generally be related to $t_c$ through this dependence. In the opposite case, where $t_c\gg T$ the exponential growth of the IR\footnote{Here we use the jargon of Effective Field Theory and call \textit{infrared} modes (IR) those of (very/sufficiently) low momenta, and \textit{ultraviolet} ones (UV) those of (very/sufficiently) high momenta.} modes does not show up until the background has changed substantially. This situation allows to make physically sensible predictions related to the behavior of UV modes in this background. Indeed, perturbation theory will provide physically meaningful predictions for all modes satisfying $\omega^{-1}\ll T$, since those modes will be insensitive to background evolution if the time intervals considered are short enough. Regarding the stability properties of the background, note that the tachyonic instability was derived assuming a constant background, and the background changes significantly much before the exponential growth of the tachyonic instability becomes relevant. Hence, in this case one cannot conclude that the background is unstable from an analysis where time-derivatives of the background of order $\mathcal{O}(T^{-n})$ have been neglected, and the problem must be studied taking into account the full dynamics of the background to truly assess its stability properties.

\begin{figure}
    \centering
    \includegraphics[scale=0.7]{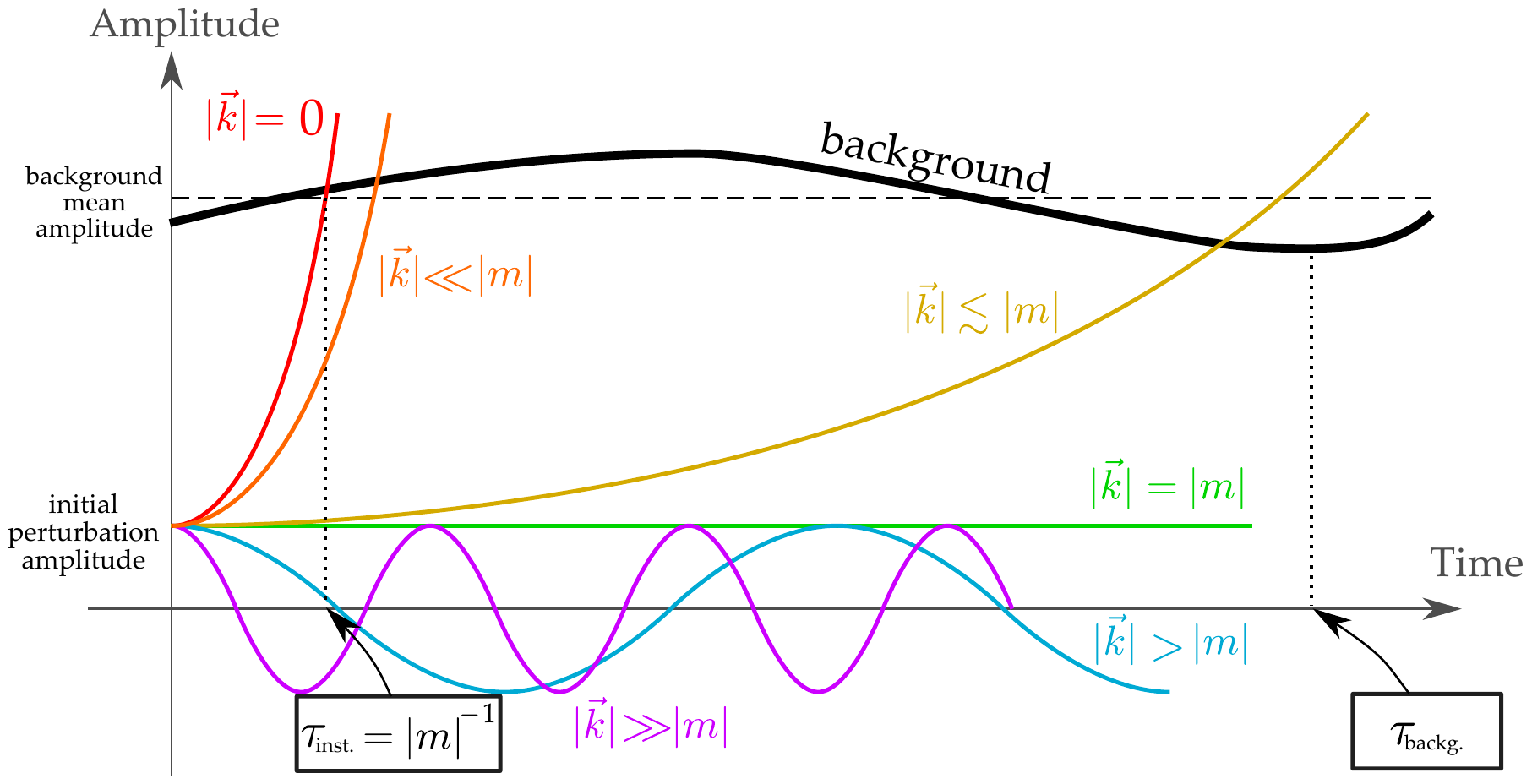}
    \caption{Orientative plot of the modes of a system with a tachyonic instability characterized by $k_{low}=m$ and with with $T>>t_c$. There is a fastest growing mode, namely that wit $|\vec{k}|=0$, that grows as $\sim \eN^{m t}$. The perturbative approach can be employed  for $t<<m^{-1}$. [Source: \cite{JCAthesis}, reproduced with the permission of the author].}
    \label{fig:my_label2}
\end{figure}

As an example of a system where an instability is crucial to model correctly the occurring physics, let us discuss the dynamics of a nearly homogeneous gas cloud with density and pressure $\bar{\rho}$ and $\bar{P}$. Now, let us allow for the gas to have (arbitrarily) small density fluctuations, so that some regions are a little bit denser than others. The equations governing the system are conservation of matter and momentum (continuity equations) and Poisson's equation linking gravitational potential to mass density of the perfect fluid. Assuming a linear relation between density and pressure perturbations in the cloud $\delta P=c_s^2\delta\rho$ (namely they are adiabatic), density perturbations are described by
\begin{equation}
    \partial_t^2{\delta\rho}-c_s^2\Delta\delta\rho-4\pi G\bar\rho\delta\rho=0
\end{equation}
where $G$ is Newton's constant. Here we can make the identification $a=1$, $b=c_s^2$ and $\mu=-4\pi G \bar{\rho}$. Therefore we will have a tachyonic instability for modes satisfying $|\vec{k}|<k_J:=\sqrt{4\pi G}/c_s$. This translates into an exponential growth of density fluctuations involving scales larger than the Jeans' length $\lambda_J:= 2\pi k_J^{-1}$. Hence, in a nearly homogeneous gas cloud, no matter how small are the density deviations from the average, if they occur over a large enough distance (the Jeans length), the density perturbations will grow quickly, implying a collapse of the cloud. Of course, this situation will in general stop at some point, when the fluctuations are big enough, fusion comes into play, changing the composition of the fluid and spoiling the linear relation between density and pressure perturbations. In that case the pressure of the collapsed fluid stabilizes the tachyonic growth leading to the formation of a star. We thus see that this instability is crucial to predict the formation of stars out of interstellar gas clouds and, indeed, we should view this instability as a transient regime from one unstable background $\bar{\rho}=\text{constant}$ (a homogeneous gas cloud) to another stable\footnote{As long as fusion is able to produce the necessary pressure to avoid further collapse!} background where matter has collapsed into a star possibly leaving some traces of the old gas orbiting around it. This example emphasizes the idea stated before that (tachyonic) instabilities are not always problenmatic, but they can be rather necessary to explain observed transient regimes in certain physical systems.

\subsubsection{Gradient/Laplacian instability: problems with UV modes}
\vspace{.3cm}
\noindent{$a>0$ and $b<0$ \hspace{.5cm}  or \hspace{.5cm} $a<0$ and $b>0$}

In this case we see that for UV modes $\omega$ becomes purely imaginary, thus triggering exponential growth/decay. Unlike the tachyonic case, for the gradient/Laplacian instabilities the growth rate is arbitrarily fast, and therefore backgrounds supporting gradient/Laplacian instabilities are always unstable and therefore perturbation theory on top of them is physically meaningless. 

\begin{figure}
    \centering
        \includegraphics[scale=0.7]{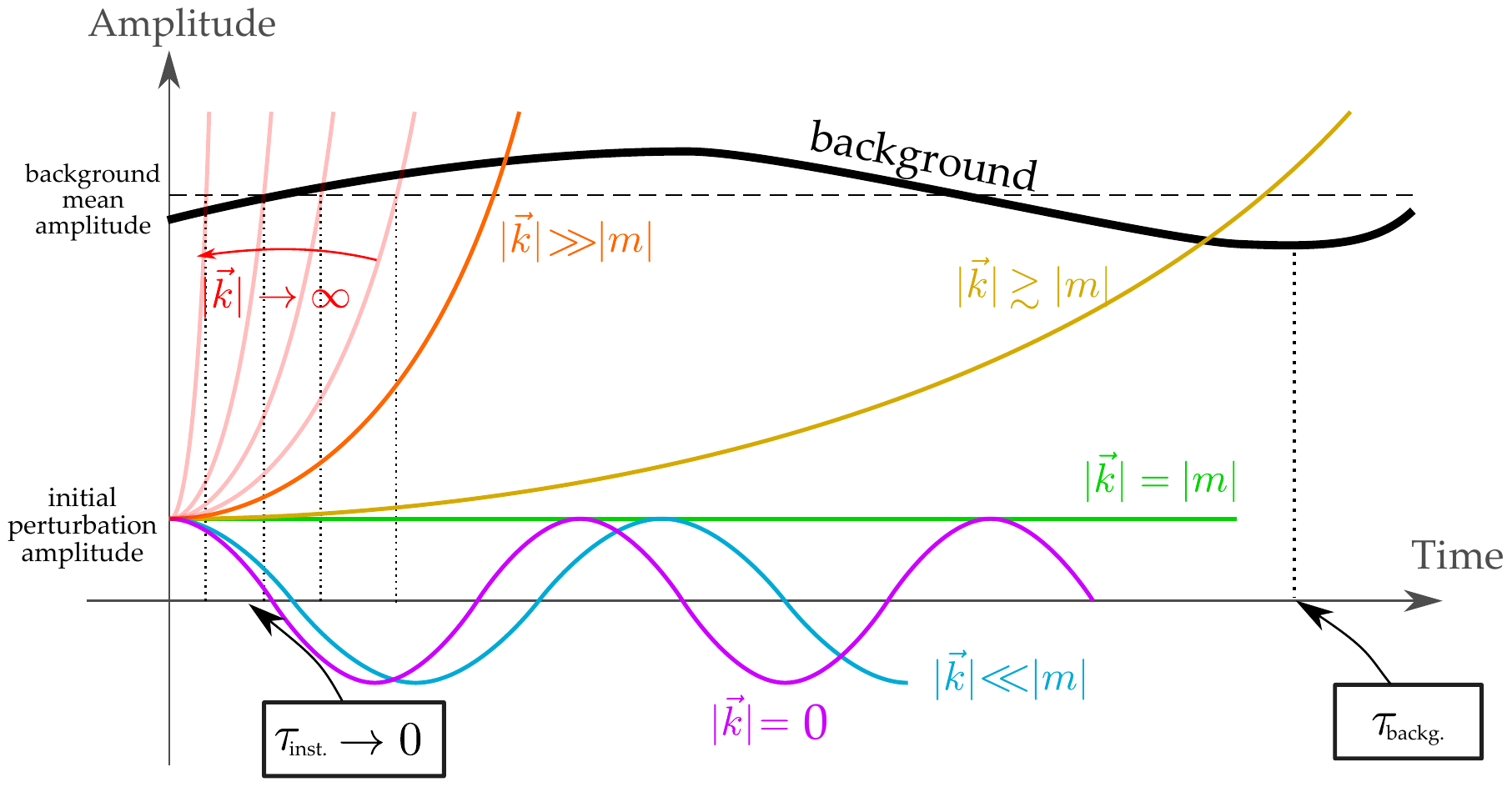}
    \caption{Orientative plot of the modes of a system with a gradient instability (for the case $\mu/b>0$). There are arbitrarily fast growing modes that spoil the perturbative approach at arbitrarily short times. [Source: \cite{JCAthesis}, reproduced with the permission of the author].}
    \label{fig:my_label}
\end{figure}

As a remark, note the in the case where $a\mu>0$, IR modes satisfying $|\vec{k}|<\sqrt{|\mu/b|}$ do not develop instabilities. Hence one could in principle devise an effective theory with a cutoff below $|\vec{k}|=\sqrt{|\mu/b|}$. Although such low energy theory would be stable if analyzed in isolation, we must recall that the background over which the effective theory is built is destabilized arbitrarily fast by the UV modes which are left outside of the effective theory. Thus, even if the low energy effective theory makes physical sense, it does not make physical sense to think of it as the low-energy limit of a theory with a gradient/Laplacian instability. Rather, if there is any reason to take this low energy effective theory seriously from a physical perspective, it should be understood as the low-energy theory of some unknown UV theory free of these instabilities.

\vspace{.3cm}
\subsubsection{Ghostly instabilities}
\noindent{$a<0$ and $b<0$}

Lastly, for negative $a$ and $b$ coefficients, we see that UV modes are always stable, whatever the sign of $\mu$, and for $\mu>0$ a tachyonic instability will appear in the IR. Thus, we see that, for a single field theory like \eqref{eq:EffectiveScalarLag}, perturbations look stable for $\mu<0$  and propagate a tachyonic instability for $\mu>0$. Nevertheless, note that the associated Hamiltonian\footnote{The Hamiltonian is $T_{00}$ with $\dot{\phi}$ written as $\pi/a$, where $\pi$ is the momentum conjugated to $\phi$.} is not bounded below (in the direction of increasing momenta), so that it does not have a ground state. Let us explore the consequences of this result both for classical an quantum theories. 

Starting with classical theories, note that for the $\mu<0$ case perturbations are oscillatory and for the $\mu>0$ case there is a tachyonic instability. However, in both cases we have a Hamiltonian unbounded from below in the direction of increasing momentum. If the field is not interacting, conservation of energy will ensure that the system does not fall to such states.\footnote{
    Note that the background is time dependent, so energy conservation is only approximate. However, the breaking terms will be of order $O(T^{-1})$, so whenever they become relevant, our analysis of stability is no longer valid. Therefore energy is conserved to the precision to which our analysis is valid. As well, there might be other conserved quantities that delimit the phase space region that is dynamically reachable from given initial conditions. If such region does not include arbitrarily high momentum states, then the energy of the system will also be dynamically bounded.}
In an interacting field theory, however, the field could decay to arbitrarily low (negative) energy states by exciting other fields to arbitrarily high energy states in what is usually called a \textit{ghost} (or runaway) instability. Though the energy of the whole system is conserved, this can lead to strange behaviors. For instance, considering a gas of interacting particles through elastic scattering, by having a particle with such a runaway instability, we can see that though the energy of the gas will be kept constant, the temperature and pressure of the gas will grow exponentially with time due to the interactions with the runaway particle, which increase the mean velocity of the particles in the gas. Of course, for this type of instability to become manifest, the runaway particle should be allowed to access regions of phase space with arbitrarily high momentum. Was there any set of symmetries that forbids this region by enforcing conservation laws, the dynamics of such system would avoid this type of instability.

In a quantum theory, again, if the ghostly field is not interacting, there is no problem due to conservation of energy. However, if it is interacting with other fields, it renders any state $|0\rangle$ that we define as the vacuum of the theory unstable. To see this note that, since $|0\rangle$ is not a ground state of the Hamiltonian (it is unbounded below), the process\footnote{
    Here $A$ is any field interacting with $\phi$. Note that in any realistic theory $\phi$ will interact at least gravitationally due to the universality of the gravitational interaction, and this process can be mediated at least by graviton exchange.}
$|0\rangle\rightarrow\varphi\varphi+A\bar{A}$ is always allowed by conservation of energy for arbitrarily high momentum of the outgoing particles. Hence, the accessible phase space volume for the final particles is infinite, making the phase space integral entering the decay rate diverge, so that the decay of the vacuum is instantaneous. If this occurs within an effective theory, the cutoff of the theory $\Lambda$ delimits the available volume of phase space that should be taken into account since the theory is not valid to arbitrarily high energies. In that case, the rate at which a tree-level process mediated by an interaction with a coupling $\alpha$ with mass dimension $d$ occurs will be of order $\alpha^2\Lambda^{4+2d}$. Given that gravity interacts universally, by considering the process $|0\rangle\rightarrow\varphi\varphi+\gamma\gamma$ mediated by gravity, which leads to a creation rate of  $\textrm{M}_\textrm{P}^{-4}\Lambda^8$, it is possible to infer an upper limit for the cutoff scale of any effective theory containing ghostly degrees of freedom from the observed spectrum of gamma rays coming from outer space of $\Lambda \lesssim 3 \,\textrm{MeV}$ \cite{ClineJeon2004}. Some authors have also attempted to provide quantization schemes in which the presence of ghostly degrees of freedom does not lead to physical inconsistencies by allowing for antilinear Hamiltonian operators, but it is still a matter of debate among the community \cite{BenderMannheim2008, ChenFasiello2013, Smilga2017}. Also the possibility that nonperturbative solutions in theories with ghosts stabilize the theory by forming a ghost condensate has been suggested \cite{Arkani-Hamed:2003pdi}.

\subsection{Generalization for N degrees of freedom}

Let us now generalize the identification of different types of instabilities explored above explored above to a theory with several degrees of freedom. As you might have noticed at this point, the presence or absence of instabilities for any degree of freedom of the theory has to do exclusively with the coefficients that define the differential operator which controls their dynamics. If we have $N$ propagating\footnote{
    Here we want to emphasize the word propagating, for reasons explained below.}
degrees of freedom, no matter how they transform under different transformation groups, we can always build an $N$-tuple $\phi^I$ out of them with $I=1,...,N$. By doing so, we will be able to express the linear part of any Lagrangian describing the propagation of $N$ degrees of freedom on top of an isotropic background as
\begin{equation}\label{eq:LagSeveralDOF}
\mathcal{L}=\frac{1}{2}\big(a_{IJ} \dot\varphi^I\dot\varphi^J- b_{IJ} (\partial^i\varphi^I)(\partial_i\varphi^J)-\mu_{IJ} \varphi^I \varphi^J\big),
\end{equation}
independently of how each of the components of $\phi^I$ transforms under different transformation groups. The field equations derived from the above Lagrangian are
\begin{equation}\label{eq:FieldEqMultiple}
    a_{(IJ)}\ddot\varphi^J-b_{(IJ)}\Delta\varphi^J-\mu_{(IJ)}\varphi^J=\mathcal{O}(T^{-1},L^{-1})
\end{equation}

The matrices  $a_{IJ}$ and $\mu_{IJ}$ are called kinetic and mass matrix, and we will call $b_{IJ}$ the gradient matrix.\footnote{
   We use the usual notation for total (anti)symmetrization of an object with indices where $2A_{[ij]}=A_{ij}-A_{ji}$, $2A_{(ij)}=A_{ij}+A_{ji}$, and the same applies for $n$ indices with the appropriate $n!$ factor on the left hand side.}
They determine the exact differential operator defining each of the PDEs that governs the propagation of each degree of freedom. Though redundant, the emphasis on propagating was put because the term degree of freedom is sometimes used in the literature to call any field appearing in the Lagrangian. However, if the Lagrangian is that of a constrained theory, some of the fields appearing will not propagate. In order to carry on this analysis, one must first solve the constraints and write the Lagrangian only in terms of truly propagating fields. For example, in a Proca theory, usually formulated in terms of four fields $A_{\mu}$ that transform as a vector under Lorentz transformations, $A_0$ is non-dynamical, and there is a constraint equation yielding $A_0=\partial_i \dot{A}^{i}$. One needs to integrate out the non-dynamical $A_0$ field by using such constraint in the Lagrangian before carrying the stability analysis as outlined here.

To assess stability of the system defined by \eqref{eq:LagSeveralDOF}, we just have to look at the positiveness properties of the kinetic, gradient and mass matrices. Negative eigenvalues will be tied to instabilities in the same way than for the single-field case. Particularly, if the different matrices commute, they can be diagonalized together, and one can look at the stability properties of each field-eigenvector by analogy to the single-field case. On the other hand, performing the stability analysis in a case in which they cannot be simultaneously diagonalized (or diagonalized at all) requires a detailed study of the particular system.

As a remark, let us discuss what are these field-eigenvectors associated to the different matrices and how they are related to the original fields. To do that, we should introduce the concept of \textit{field redefinition}. Much as we can do coordinate transformations in classical point-particle mechanics in order to work with any desired set of phase space coordinates, we can also do this kind of transformations in field theory, and work with different field variables. Different field variables will generally have different physical meaning, but all will encode the same information about the system if the transformation from one to the other is well defined.\footnote{Mainly invertible and with some required number of invertible derivatives.} Having said this, note that if an eigenvectors of one of these matrices is $\alpha$, with components $\alpha_I$ in the basis in which the matrix is written, it will define a linear combination of the original fields by $\Psi_\alpha=\alpha_I\phi^I$, with each eigenvector yielding a different linear combination of the original fields. Thus these field-eigenvectors are nothing but a linear field redefinition of the original fields which takes corresponding matrix to diagonal form. Using these field space variables, if they exist, the question of what degrees of freedom of the system are unstable and what kind of instability do they develop becomes straightforward by analogy with the single-field case.

\subsection{Strongly coupled backgrounds and their unstable nature}

Another type of frequent instability that arises in field theories is the so-called \textit{strong coupling} instability. Unlike the ones discussed above, this type of instability cannot be explicitly seen at linear order in perturbation theory, although one might already suspect its presence from the linear analysis. This is so because this instability occurs when some degree of freedom of a theory does not propagate on top of a particular vacuum because the kinetic term of such mode vanishes on such background. So, in short, one would expect this kind of instability to occur when there is a particular background where some of the eigenvalues of the kinetic matrix vanish, but they do not vanish in some field configurations that are arbitrarily close to such background. Thus, naively, if we canonically normalize the kinetic term of the propagating degrees of freedom,\footnote{Think of it roughly as multiplying \eqref{eq:FieldEqMultiple} by $(a_{(IJ)})^{-1}$ on the left.} we expect that the coefficients of the rest of the terms in the Lagrangian featuring the strongly coupled degree of freedom diverge.

To see this, consider a theory depending on a set of fields $\{\phi^I\}$, and a background $\{\Phi^I_0\}$ around which the perturbations of one of the fields (that we will call $\psi$) do not propagate. Consider also that we approach such a background with a curve in solution space  $\{\phi^I_0(\lambda)\}$ with $\lambda\in[0,1]$, such that  for $\lambda<1$ there is no strong coupling (i.e., $\psi$ propagates) and  $\{\phi^I_0(1)=\Phi^I_0\}$. If we expand around any point of this curve with $\lambda<1$, we can extract from the Lagrangian the kinetic term of the perturbations $\psi$ and it will have the form:
\begin{equation}
    \mathcal{L}_\lambda=\frac{1}{2}F(\phi^I_0(\lambda))\dot{\psi}^2 + \mathcal{L}_\text{rest}\,.
\end{equation}
This Lagrangian must satisfy (by construction) that $F(\phi^I_0(1))=F(\Phi^I_0)=0$. Notice that if we canonically normalize the perturbation,
\begin{equation}
    \mathcal{L}_\lambda|_\text{canon.}=\frac{1}{2}\dot{\psi}^2 + \frac{1}{F(\phi^I_0(\lambda))}\mathcal{L}_\text{rest} \,,
\end{equation}
all the interactions (non-kinetic couplings) of $\psi$ will become infinite around $\{\Phi^I_0\}$
\begin{equation}
    \lim_{\lambda\to 1}\frac{1}{F(\phi^I_0(\lambda))}\mathcal{L}_\text{rest}= \infty\,.
\end{equation}
This is the reason why it is said to be strongly coupled.

In general, we expect this to happen when we have a theory that propagates $N$ degrees of freedom in the general case but linear perturbations on top of a particular vacuum of the theory describe $M<N$ propagating modes. In that case, it may (or may not) occur that we are able to see explicitly how the strongly coupled modes are present by going at higher order in perturbations. However, there is no general rule to know at which order will those appear. This discontinuity in the number of degrees of freedom when going to higher order perturbations signals the breakdown of the perturbative expansion, which then yields meaningless results on top of a strongly coupled background. This occurs because the system of differential equations describing time evolution reduces its dimensionality around such a vacuum, namely the range of the kinetic matrix decreases abruptly.  From the phase space perspective, strongly coupled backgrounds can typically be linked to some geometrical singularity in phase space. Note that, though this instability can be guessed when one knows the number of propagating degrees of freedom of a theory in the general case, it is very hard to foresee in cases where this information is not available since, in that situation, one does not know whether perturbations around a particular background describe less degrees of freedom than around other backgrounds. Thus, this instability will generally stay invisible unless we come across the situation where we know the number of propagating degrees of freedom on top of two different backgrounds does not coincide, or we find a change in the number of degrees of freedom at different orders in perturbation theory.

\newpage
\begin{center}
  \textbf{LIST OF PROBLEMS FOR LECTURE 1}   
\end{center}
\addcontentsline{toc}{section}{List of problems for Lecture 1}

\noindent{}\textbf{Problem 1.} \textit{Bounded and unbounded growth}
\vspace{.3cm}

\begin{itemize}
    \item[a)] Show that indeed the perturbations for the free scalar field are bounded, namely, that for any initial perturbation $\phi$ there exists a constant $C$ such that $|\varphi|<C$ at all times. 
    \item[b)] Show also that, for a system with a gradient instability, given initial amplitude $A$, and an arbitrarily small time interval $\Delta t$, there are modes with unbounded growth within that time interval. Show also that is not true for a system with a tachyonic instability.
\end{itemize}

\vspace{1cm}

\noindent{}\textbf{Problem 2.} \textit{Jean's instability}
\vspace{.3cm}

\noindent{}For a self gravitating cloud of dust of density $\rho$ and pressure $P$, the mass and momentum conservation equations are 
    \begin{equation*}
        \partial_t\rho+\partial_i(\rho u^i)=0 \qquad\text{and}\qquad\partial_t u^j+ u^i\partial_i u^j+\rho^{-1}\partial^jP+\partial^j\Phi=0
    \end{equation*}
where $u^i$ is the velocity vector of a fluid element, $\Phi$ is the gravitational potential satisfying the Poisson equation $\Delta\Phi=\kappa\rho$, and indices are raised and lowered with $\delta_{ij}$. Assuming a steady initial state $\bar u^i=0$ with constant density and pressure $\bar\rho$ and $\bar P$:
\begin{enumerate}[label=\alph*)]
    \item Derive the equation followed by adiabatic perturbations $\delta P=c_s^2\delta\rho$. 
    \item For the case where gravitation is turned off, $\kappa=0$, describe the behavior of adiabatic density perturbations. Is the initial configuration of the system stable?
    \item Repeat the analysis taking into account the gravitational interaction. What qualitative differences do you find and what is the physical interpretation for them?
    \item If one takes into account the expansion of the universe, and define the fractional density perturbation $\delta:=\delta\rho/\bar\rho$, the equation that you have found becomes 
    \begin{equation}
        \partial_t^2{\delta}+3H\partial_t\delta-c_s^2\Delta\delta-\kappa\rho\delta=0
    \end{equation}
    For constant positive expansion rate $H>0$, compute the general solution to this equation. Discuss what is the role of the expansion rate and whether it changes anything about the stability of the perturbations from a qualitative point of view.
\end{enumerate}

\vspace{1cm}
\noindent{}\textbf{Problem 3.} \textit{Non-canonical kinetic and mass terms: Uncovering instabilities through field redefinitions.}
\vspace{.3cm}

\noindent{}The following Lagrangian density describes a field theory with two fields $\phi$ and $\psi$ propagating around a particular background defined by the coefficients $a$, $b$ and $c$.
    \begin{equation*}
        \mathcal{L}=a\,\partial^\mu\phi\partial_\mu\phi+b\,\partial^\mu\psi\partial_\mu\psi+c\,\partial^\mu\phi\partial_\mu\psi + d\, \phi\psi
    \end{equation*}
\begin{enumerate}[label=\alph*)]
    \item find the kinetic, gradient and mass matrices for the theory.
    \item For $b=0$, show that there is a ghost degree of freedom.
    \item Show that whenever $4ab\neq c$ the theory propagates 2 degrees of freedom. Derive the condition/s that must be satisfied by the coefficients $a,b,c$ for the theory to be free of ghosts and/or gradient instabilities. 
    \item What relations must the coefficients satisfy in order for the background to potentially suffer from a strong coupling instability? If this condition is satisfied for this background, in what cases will we have this instabilities?
    \item Find the field redefinition that diagonalizes the mass matrix (i.e. the fields that are mass eigenstates) and show that there is always a tachyonic instability unless the fields are massless.
\end{enumerate}

\newpage

\lecture{Ostrogradski theorem and ghosts}
\addcontentsline{toc}{section}{Ostrogradski theorem and ghosts}

In the previous lecture, we studied a variety of problems that can arise in field theory around particular backgrounds. The Lagrangians we considered contain, as usual, at most first-order derivatives of the fields. In the present lecture, we focus on the case of Lagrangians containing higher-than-first-order derivatives. Usually, these derivatives imply the need of additional initial conditions to determine the evolution of the system, showing that the fields of the theory hide extra degrees of freedom (which can be extracted with appropriate field redefinitions). As we will see, when these extra degrees of freedom are present, ghosts inevitably appear and generically propagate around the solutions of the theory, making it invalid for physical purposes.

In this lecture, we will start by quickly revising the `standard' Hamiltonian analysis for field theories and extending it to theories with higher-than-first-order derivatives.

\subsection{Some preliminary ideas}

Consider a theory (in flat space) depending on some family of fields $\phi^I$ and up to first-order derivatives of them,
\begin{equation}
    S[\phi^I] = \int \mathcal{L}(\phi^I,\ \partial_i\phi^I,\  \dot{\phi}^I) \dd^4x = \int L[\phi^I, \dot{\phi}^I] \dd t \,,\label{eq:Lfdf}
\end{equation}
where we are denoting $\dot{X}:= \partial_0X$ and the indices $i,j...$ cover the spatial directions. Here we have introduced the Lagrangian functional $L[\phi^I, \dot{\phi}^I]$,
\begin{equation}
    L[\phi^I, \dot{\phi}^I] := \int \mathcal{L}(\phi^I,\  \partial_i\phi^I,\  \dot{\phi}^I) \dd^3x \,.
\end{equation}
The Euler-Lagrange equations are
\begin{equation}
    0=\frac{\delta S}{\delta \phi^I}= \frac{\delta L}{\delta \phi^I}-\frac{\partial}{\partial t}\frac{\delta L}{\delta \dot{\phi^I}}=\frac{\partial \mathcal{L}}{\partial \phi^I}-\partial_\mu \frac{\partial \mathcal{L}}{\partial (\partial_\mu\phi^I)},
\end{equation}
and, after applying the chain rule in the last term, we get
\begin{equation}
    0= (\text{independent of}\,\{\ddot{\phi}^J\})_I -\mathcal{K}_{IJ}\ddot{\phi}^J\,.
\end{equation}
Here, we defined
\begin{equation}
    \mathcal{K}_{IJ}:=\frac{\partial^2  \mathcal{L}}{\partial \dot{\phi}^I \partial \dot{\phi}^J},
\end{equation}
called the \emph{Hessian matrix}. For the Lagrangians \eqref{eq:Lfdf} that we normally use in physics, the Hessian corresponds with the kinetic matrix (maybe up to normalization). In addition, we assume that the theory fulfills the \emph{non-degeneracy condition},\footnote{We will give a more general definition for `Hessian' and `non-degeneracy' later, which will be valid for also for Lagrangians with higher-than-first-order derivatives.} 
\begin{equation}
    \det(\mathcal{K}_{IJ})=\det\left(\frac{\partial^2  \mathcal{L}}{\partial \dot{\phi}^I\partial\dot{\phi}^J}\right) \neq 0\,,\label{eq:nondeg0}
\end{equation}
because this allows to re-express the Euler-Lagrange equation as
\begin{equation}
    \ddot{\phi}^J= (\mathcal{K}^{-1})^{JI}\times (\text{independent of}\,\{\ddot{\phi}^J\})_I\,.
\end{equation}
which allows to solve for the accelerations in terms of the fields and the velocities:
\begin{equation}
   \phi^I(t, \vec{x})= f^I(t, u(\vec{x}),\dot{u}(\vec{x}))\,.
\end{equation}
where $u$ and $\dot{u}$ represent the initial conditions for the field and its velocity, respectively, in a spacelike hypersurface.

In flat space, for this theories, we can perform the Hamiltonian construction as follows. First introduce the canonical variables:
\begin{equation}
    q^I :=\phi^I\,,\qquad \pi_I:= \frac{\delta L}{\delta \dot{\phi}^I} = \frac{\partial \mathcal{L}}{\partial \dot{\phi}^I}\,.
\end{equation}
If the Lagrangian is non-degenerate we can solve for the velocities in the definition of the momenta, $\dot{q}^I=\dot{q}^I(\pi_J)$, and we are able to introduce the Hamiltonian functional:\footnote{Notice that if the condition \eqref{eq:nondeg0} is not fulfilled, then the velocity cannot be solved in terms of the momentum. This is signaling the presence of constraints that should be appropriately treated (e.g., via the Dirac procedure).}
\begin{equation}
    H[q^I, \pi_I] := \int \dot{q}^I|_{q,\pi}\, \pi_I\ \dd^3x - L|_{q,\pi} = \int \big(\dot{q}^I|_{q,\pi}\, \pi_I - \mathcal{L}|_{q,\pi}\big) \dd^3x\,,
\end{equation}
where $X|_{q,\pi}$ indicates that $X$ should be written in terms of the canonical variables. The dynamics is then determined by the Hamilton equations:
\begin{equation}
    \dot{q}^I=\frac{\delta H}{\delta \pi_I},\qquad \dot{\pi}_I=-\frac{\delta H}{\delta \phi^I}\,.
\end{equation}
These equations require initial conditions for $q^I$ and for $\pi_I$, which are essentially equivalent to those of $\phi^I$ and $\dot{\phi}^I$ in Lagrangian formulation. It is worth remarking that a \emph{true} (or \emph{propagating}) degree of freedom requires is a pair of initial conditions, one for itself and another one for its velocity or momentum. 

In the next section we are going to extend this procedures to Lagrangians with higher-than-first-order derivatives. As we will see, in such a case, a single field $\phi$ could lead to two (or more) degrees of freedom.

\subsection{Lagrangian with higher order derivatives}

It is clear, by construction, that the formalism of the previous section is not valid for Lagrangians containing higher-than-first-order derivative (except when they can be eliminated after integration by parts). The generalization of the previous formalism was done in 1850 by Ostrogradski in his work \cite{Ostrogradski1850}. To deal with it, we first introduce the following concepts
\begin{boxmathdark}
\noindent\textbf{Definition. Hessian}\\
The \emph{Hessian of a Lagrangian} $L$ is the matrix of second variations of $L$ with respect to the highest time derivatives of the fields (after removing the global Dirac delta). 
\end{boxmathdark}

This definition might seem a bit obscure,\footnote{
   The issue of the delta is due to the fact that second variations of integral functionals are proportional to the Dirac delta or spatial derivatives of it. So, after computing the matrix of second variations, one should multiply by a test function and integrate to remove the delta. The result is matrix which contains combinations of fields or operators. Such a matrix is the Hessian.}
but for the cases we are going to deal with in these lectures, the Hessian can be written in a more familiar way as the matrix of second partial derivatives of $\mathcal{L}$ with respect to the highest time derivatives of the fields.

\begin{boxmathdark}
\noindent\textbf{Definition. Non-degenerate Lagrangians}\\
A Lagrangian is said to be \emph{(non-)degenerate} if the Hessian has (non-)vanishing determinant.
\vspace{-0.15cm}
\end{boxmathdark}

Degeneracy is connected with the fact that some of the highest order derivatives can be eliminated for instance after integration by parts or by redefining the fields appropriately. Let us see a very simple example of a degenerate with first derivatives:
\begin{boxmathclear}
\noindent{\bf Example. Trivial degenerate Lagrangian}

For instance, consider the theory,
\begin{equation}
    S[\phi] = \int (\phi \dot{\phi}  - \phi^2) \dd^4x.
\end{equation}
It only depends up to first-order derivatives of one field, so we do not expect more than one degree of freedom.

Let us first check that it is degenerate:
\begin{equation}
    \frac{\delta L}{\delta \dot{\phi}} = \frac{\partial \mathcal{L}}{\partial \dot{\phi}} =  \phi  \qquad \Rightarrow\qquad \frac{\delta^2 L}{\delta \dot{\phi}^2} = 0.
\end{equation}
This implies that we cannot solve for the field in terms of `position+velocity' initial conditions (as it should be for one degree of freedom)
\begin{equation}
   \phi(t, \vec{x}) \neq f(t, u(\vec{x}),\dot{u}(\vec{x}))\,.
\end{equation}
In fact, one can extract appropriate boundary terms in the time direction to eliminate them:
\begin{equation}
    S[\phi] = \int \left[ \partial_0 \left(\frac{1}{2}\phi^2\right) - \phi^2 \right] \dd^4x = \int ( - \phi^2 ) \dd^4x + \text{boundary term}
\end{equation}
So this theory is clearly non-dynamical. The only solution is $\phi=0$ which has no non-trivial time evolution. So this theory propagates no degrees of freedom.

Indeed, if we go to the Hamiltonian picture (we call $q:=\phi$) the degeneracy leads to a constraint when computing the momenta (primary constraint):
\begin{equation}
    \left(\frac{\delta L}{\delta \dot{\phi}}=:\right)\pi= q \qquad \Rightarrow \qquad \mathcal{C}:=\int \dd^3x (\pi - q)\,.
\end{equation}
The Hamiltonian of the theory is
\begin{equation}
    H[q,\pi]=- \int q^2 \dd^3 x + \lambda \mathcal{C}(q,\pi)
\end{equation}
where $\lambda$ is a Lagrange multiplier. The time evolution of the constraint $\mathcal{C}$ leads to another constraint $\mathcal{D}=q$, which finishes the Dirac algorithm. It can be checked that the Poisson bracket of these two constraints is a constant number. Therefore, we have two second-class constraints, and this reduces in one unit the number of degrees of freedom. So, in conclusion, the theory propagates 0 degrees of freedom (i.e., there are no dynamical variables in it), as we showed previously in the Lagrangian formulation.
\end{boxmathclear}

Consider now a field theory depending on just one scalar field and up to second-order time derivatives:
\begin{equation}
    L[\phi, \dot{\phi}, \ddot{\phi}]=\int \mathcal{L}(\phi,\  \partial_i\phi,\  \dot{\phi},\  \partial_i\partial_j\phi,\  \partial_i\dot{\phi},\  \ddot{\phi})\ \dd^3 x\label{eq:LcalL}
\end{equation}
As we already said, this Lagrangian functional is assumed to be non-degenerate. In this case, this is true if and only if
\begin{equation}
  \frac{\delta^2 L}{\delta \ddot{\phi}^2}\neq0,
\end{equation}
which in this case is equivalent to\footnote{
   The second variations of $L$ are distributional objects proportional to the Dirac delta or spatial derivatives of it. In the cases we will study in these lectures no derivatives of the delta appear in the Hessian. In these cases, one can multiply by a test function and integrate to eliminate the delta, and the result is the matrix of second partial derivatives of $\mathcal{L}$ (see e.g. \eqref{eq:checkdegLBfi}).}
\begin{equation}
\frac{\partial^2 \mathcal{L}}{\partial \ddot{\phi}^2}\neq0 \,.
\end{equation}

Similarly as in the previous section, now the Euler-Lagrange equation,
\begin{equation}\label{eq:ELeqOstro}
  0=\frac{\delta S}{\delta \phi} \ = \ \frac{\delta L}{\delta \phi}-\frac{\partial}{\partial t} \frac{\delta L}{\delta \dot{\phi}}+\frac{\partial^2}{\partial t^2}\frac{\delta L}{\delta \ddot{\phi}}\ =\ \frac{\partial \mathcal{L}}{\partial \phi}-\partial_\mu \frac{\partial \mathcal{L}}{\partial (\partial_\mu \phi)}+\partial_\mu\partial_\nu\frac{\partial \mathcal{L}}{\partial (\partial_\mu\partial_\nu\phi)}\qquad 
\end{equation}
can be written, after using the chain rule, as
\begin{equation}
   \ddddot{\phi}=  \left(\frac{\partial^2 \mathcal{L}}{\partial \ddot{\phi}^2}\right)^{-1}\times (\text{something independent of}\,\ddddot{\phi})
\end{equation}
which locally admits solutions of the type
\begin{equation}
   \phi(t,\vec{x})=f\Big(t,\ u(\vec{x}),\ \dot{u}(\vec{x}),\  \ddot{u}(\vec{x}),\ \dddot{u}(\vec{x})\Big)\,.
\end{equation}
As in the previous section, $\{u,\dot{u},\ddot{u},\dddot{u}\}$ are the initial conditions for $\{\phi,\dot{\phi},\ddot{\phi},\dddot{\phi}\}$, respectively, in a spacelike hypersurface. Observe, how the presence of one additional derivative order in the Lagrangian requires two additional initial conditions, $\ddot{u}$ and $\dddot{u}$. Each pair of initial conditions correspond to a propagating degree of freedom, so here we have an example of a scalar field that contains two degrees of freedom. The extra one is hidden in the fact that there are higher than first order derivatives in the (non-degenerate) Lagrangian.

Let us remark one more time that the non-degeneracy condition is what allows us to solve for the highest order derivative terms of the equations of motion, which are of the order $2\times$the highest order of the Lagrangian. For instance, in the previous case, in which the Lagrangian depends up to second derivatives of $\phi$, the non-degeneracy permits to solve the equations for $\ddddot{\phi}$. Now we present an example of a particular Lagrangian depending on two fields, showing how all of these ideas can be straightforwardly generalized to multi-field theories:
\begin{boxmathclear}
\noindent{\bf Example. Lagrangian with two fields up to 2nd-order derivatives of one field.}

Consider the Lagrangian
\begin{equation}
    \mathcal{L}=\frac{1}{2}\ddot{\phi}^2 + \frac{1}{2}\dot{\psi}^2 + a \ddot{\phi}\dot{\psi}+\frac{1}{2}\phi^2\,.
\end{equation}
The Hessian can be explicitly read from the Lagrangian as follows:
\begin{equation}
    \mathcal{L}={\frac 12} (\ddot{\phi}, \dot{\psi}) 
  \begin{pmatrix}1 & a\\a & 1\end{pmatrix}
  \begin{pmatrix}\ddot{\phi}\\ \dot{\psi} \end{pmatrix}+\frac{1}{2}\phi^2\,.
\end{equation}
So the theory is non-degenerate, only if $a\neq\pm1$. If we compute the equations of motion, we get
\begin{equation}
    [\text{EoM}\,\phi]:\quad \ddddot{\phi}+a\dddot{\psi}+\phi=0,\qquad[\text{EoM}\,\psi]:\quad \ddot{\psi}+a\dddot{\phi}=0\,,
\end{equation}
which can be combined in
\begin{equation}
    (1-a^2)\ddddot{\phi}=-\phi,\qquad \ddot{\psi}=-a\dddot{\phi}\,.
\end{equation}
Observe that we cannot solve for $\ddddot{\phi}$ and $\ddot{\psi}$ if $a=\pm1$. In the non-degenerate case ($a\neq\pm1$), we will need 4 initial conditions for $\phi$ and 2 for $\psi$, making a total of 3 degrees of freedom, one of which were hidden in the higher-order derivatives of $\phi$.
\end{boxmathclear}

\subsection{Hamiltonian procedure for higher-order Lagrangians. The Ostrogradski theorem}

To treat the general theory $L[\phi, \dot{\phi}, \ddot{\phi}]$ (which we assumed to be non-degenerate) of the previous section in the Hamiltonian language, one can make use of the Ostrogradski procedure \cite{Ostrogradski1850}.\footnote{
    Later we will see the case of arbitrarily high time derivatives. Regarding the generalization to multi-fields, one just should apply the procedure to each of the fields of the theory.} 
As we will see, this method leads to a very important result about stability in field theory, the \emph{Ostrogradski theorem}. 

Let us start with the construction of the Hamiltonian. Firstly, one should extract the information about the extra degree of freedom (i.e., the high-order derivatives) by introducing another pair of canonical variables in the Hamiltonian formalism. The canonical positions and momenta are defined as follows:
\begin{equation}\label{eq:qpiOstro}
  q_1 := \phi,\qquad q_2 := \dot{\phi}\,,\qquad 
  \pi_1:= \frac{\delta L}{\delta \dot{\phi}}-\frac{\partial}{\partial t}\frac{\delta L}{\delta \ddot{\phi}}\,,\qquad 
  \pi_2:= \frac{\delta L}{\delta \ddot{\phi}}\,.
\end{equation}
The non-degeneracy condition implies that one can invert these definitions to write
\begin{equation}
  (\dot{q}_2=)\quad\ddot{\phi} = F(q_1, q_2, \pi_2)\,,\label{eq:soldq2}
\end{equation}
such that the following equation is fulfilled:
\begin{equation}
  \pi_2 = \frac{\delta L}{\delta \ddot{\phi}} \Big|_{\phi=q_1, \, \dot{\phi}=q_2,\, \ddot{\phi}=F}\,.
\end{equation}
Observe how the second-order derivatives $\ddot{\phi}$ become $\dot{q}_2$, i.e., we have hidden the higher-order-derivative nature of the Lagrangian by introducing a new variable. Now we can construct the Hamiltonian in the usual way, as the Legendre transformation of the Lagrangian with respect to the time derivatives of the fields (in this case $\{q_1,q_2\}$):
\begin{align}
 H[q_1,q_2,\pi_1,\pi_2] &= \int \Big[\dot{q}_1 \pi_1 + \dot{q}_2\pi_2\Big] \dd^3x - L[q_1,q_2,\dot{q}_2] \\
   {}^\eqref{eq:soldq2}&= \int \Big[q_2 \pi_1 +  F(q_1, q_2, \pi_2)\pi_2\Big] \dd^3x - L[q_1,q_2,F(q_1, q_2, \pi_2)] \\
   {}^\eqref{eq:LcalL}&= \int \Big[q_2 \pi_1 +  F(q_1, q_2, \pi_2)\pi_2- \mathcal{L}\big(q_1, \partial_iq_1,q_2,\partial_i\partial_jq_1,\partial_iq_2,F(q_1, q_2, \pi_2)\big)\Big] \dd^3x  \,.
\end{align}
Here an important thing to notice is that the momentum $\pi_1$ enters linearly, which implies that the Hamiltonian is unbounded from below in the direction of the phase space given by $\pi_1$. This is a particular case of a more general result called Ostrogradski theorem. Here we reproduce it as was stated in \cite{AokiMotohashi2020}:\footnote{See also \cite{Pons1989, GanzNoui2021}.}
\begin{boxmathdark}
\noindent \textbf{Theorem (Ostrogradski, 1850)} \\
Let a Lagrangian involve $n$-th-order finite time derivatives of variables. If $n\geq2$ and the Lagrangian is non-degenerate with respect to the highest-order derivatives, the Hamiltonian of this system linearly depends on a canonical momentum.
\end{boxmathdark}

This unboundedness implies the presence of ghosts (Ostrogradski ghosts) in the theory. Therefore, whenever we are dealing with a higher-order theory, one has to ensure that the theory is degenerate, in order to bypass the devastating consequences of this theorem.

Let us see a particular example of the previous derivation:

\begin{boxmathclear}

\noindent{\bf Example. The Ostrogradski procedure.}

Consider the following Lagrangian for a scalar field
\begin{equation}
  \mathcal{L}= {\frac 12} \partial_\mu\phi\partial^\mu\phi + {\frac \lambda 2}(\square\phi)^2 - V(\phi),\qquad \lambda \neq0.
\end{equation}
where $\square:= \eta^{\mu\nu} \partial_\mu\partial_\nu$. The corresponding Lagrangian functional is
\begin{equation}
  L[\phi, \dot{\phi}, \ddot{\phi}] = \int {\rm d}^3x \mathcal{L} =  \int {\rm d}^3x \Big[ {\frac 12} (\dot{\phi}^2 - |\vec{\nabla}\phi|^2) + {\frac \lambda 2}(\ddot{\phi} - \vec{\nabla}^2\phi)^2 - V(\phi)\big]\,,
\end{equation}
which is non-degenerate, as one can easily check:
\begin{equation}
  \det \big[\mathrm{Hess}(L)\big]=\frac{\partial^2 \mathcal{L}}{\partial \ddot{\phi}^2} = \lambda  \quad\neq0\,.
\end{equation}
We start the Ostrogradski procedure by introducing the set of canonical variables 
\begin{align}
  q_1 & := \phi\,,
  &\pi_1 &:= \frac{\delta L}{\delta \dot{\phi}}-\frac{\partial}{\partial t}\frac{\delta L}{\delta \ddot{\phi}}= \dot{\phi} - \lambda \dddot{\phi} + \lambda\vec{\nabla}^2\dot{\phi} \,,\nonumber\\
  q_2 & := \dot{\phi}\,,
  &\pi_2 &:= \frac{\delta L}{\delta \ddot{\phi}} =\lambda(\ddot{\phi} - \vec{\nabla}^2\phi)\,.
\end{align}
From the expression of $\pi_2$ we can obtain the function $F$,
\begin{equation}
  \ddot{\phi} = F(q_1,q_2,\pi_2) = {\frac 1 \lambda} \pi_2 + \vec{\nabla}^2 q_1\,.
\end{equation}
In this variables the Lagrangian density has the form
\begin{equation}
  \mathcal{L} = -{\frac 12} (|\vec{\nabla}q_1|^2-(q_2)^2 ) + \frac{1}{2\lambda}(\pi_2)^2 - V(q_1)\,.
  \end{equation}
The resulting Hamiltonian functional,
\begin{align}
   & H[q_1,q_2,\pi_1, \pi_2] \nonumber\\
   &\qquad = \int {\rm d}^3x \big[\dot{q}_1 \pi_1 + \dot{q}_2\pi_2 - \mathcal{L} \big] \nonumber\\
   &\qquad =  \int {\rm d}^3x \Big[q_2 \pi_1 + \Big(\frac{1}{2\lambda}\pi_2 + \vec{\nabla}^2 q_1\Big)\pi_2 + {\frac 12} \Big(|\vec{\nabla} q_1|^2-(q_2)^2 \Big) + V(q_1) \Big] \,,
\end{align}
is indeed linear in the momentum the direction of $\pi_1$. Actually, one can separate the coupling between $\pi$'s and $q$'s, via the coordinate transformation  $(\pi_1\to\pi_1+q_2/2,\ \pi_2\to\pi_2- \lambda\vec{\nabla}^2 q_1)$ in phase space, which leads us to:
\begin{equation}
    H[q_1,q_2,\pi_1, \pi_2] =  \int {\rm d}^3x \Big[q_2 \pi_1 + \frac{1}{2\lambda}(\pi_2)^2 + {\frac 12} \Big(|\vec{\nabla} q_1|^2-\lambda(\vec{\nabla}^2 q_1)^2\Big) + V(q_1) \Big] \,.
\end{equation}

\end{boxmathclear}

~

\noindent\textit{Some remarks}

We will now proceed to highlight a couple of aspects about the Ostrogradski Hamiltonian:
\begin{enumerate}
    \item The Hamilton equations of the Ostrogradski Hamiltonian reproduce the same dynamics as the original Lagrangian. Indeed, for the case we have been treating in this section, if we compute the four variations of the Hamiltonian:
    \begin{align}
        \frac{\delta H}{\delta \pi_1} &= q_2,\\
        \frac{\delta H}{\delta \pi_2} &= F + \pi_2 \frac{\partial F}{\partial \pi_2} - \frac{\delta L}{\delta \ddot{\phi}}\frac{\partial F}{\partial \pi_2}=F,\\
        \frac{\delta H}{\delta q_1} &= \frac{\partial F}{\partial q_1}\pi_2-\frac{\delta L}{\delta \phi}-\frac{\delta L}{\delta \ddot{\phi}}\frac{\partial F}{\partial q_1} = -\frac{\delta L}{\delta \phi},\\
        \frac{\delta H}{\delta q_2} &= \pi_1 + \frac{\partial F}{\partial q_2}\pi_2-\frac{\delta L}{\delta \dot{\phi}}-\frac{\delta L}{\delta \ddot{\phi}}\frac{\partial F}{\partial q_2} = \pi_1 - \frac{\delta L}{\delta \dot{\phi}}
    \end{align}
    one easily realizes that three of the Hamilton equations are nothing but the definitions of $q_2$, $F$ and $\pi_1$, respectively,
\begin{align}
  \dot{q}_1&= \frac{\delta H}{\delta \pi_1} &\Leftrightarrow\,\qquad\dot{q}_1&= q_2\,,\label{eq:Heom1} \\
  \dot{q}_2&= \frac{\delta H}{\delta \pi_2} &\Leftrightarrow\,\qquad\dot{q}_2&= F(q_1, q_2, \pi_2)\,,\\
  \dot{\pi}_2&=-\frac{\delta H}{\delta q_2} &\Leftrightarrow\qquad\dot{\pi}_2&= -\pi_1 + \frac{\delta L}{\delta \dot{\phi}}\label{eq:Heom3}\,,
\end{align}
whereas the remaining one is the Euler-Lagrange equation \eqref{eq:ELeqOstro}:
\begin{equation}
  \dot{\pi}_1=-\frac{\delta H}{\delta q_1} \qquad \Leftrightarrow \qquad 0= \frac{\delta L}{\delta \phi}-\dot{\pi}_1= \frac{\delta L}{\delta \phi}-\frac{\partial}{\partial t}\left(\frac{\delta L}{\delta \dot{\phi}}-\frac{\partial}{\partial t}\frac{\delta L}{\partial \ddot{\phi}}\right)\,,
\end{equation}

    \item If the theory is invariant under time translations, the conserved charge (the \emph{energy} of the system) coincides with the Ostrogradski Hamiltonian \cite{Woodard2015}.
\end{enumerate}

\noindent\textit{Generalization to arbitrarily high derivatives}

This procedure can be generalized to arbitrarily high derivatives (see e.g. \cite{Woodard2015}). For a non-degenerate Lagrangian depending up to $n-$th derivatives of some field $\phi$, one needs to introduce $n$ new fields and momenta $\{q_i,\,\pi_i\}_{i=1}^{n}$ defined as
\begin{equation}
    q_i:= \phi^{(i-1)},\,\qquad \pi_i:= \sum_{j=i}^{n} \left(-\frac{\partial}{\partial t}\right)^{j-i}\frac{\delta L}{\delta \phi^{(j)}}\,,
\end{equation}
for instance 
\begin{equation}
    q_1:= \phi,\qquad 
    q_2 := \dot{\phi},\qquad 
    q_3 := \ddot{\phi},\qquad...\qquad 
    q_n := \phi^{(n-1)},
\end{equation}
\begin{equation}
    \pi_n:= \frac{\delta L}{\delta \phi^{(n)}},\quad 
    \pi_{n-1} := \frac{\delta L}{\delta  \phi^{(n-1)}}-\frac{\partial}{\partial t}\frac{\delta L}{\delta  \phi^{(n)}},\quad 
    \pi_{n-2} := \frac{\delta L}{\delta  \phi^{(n-2)}}-\frac{\partial}{\partial t}\frac{\delta L}{\delta  \phi^{(n-1)}}+\frac{\partial^2}{\partial t^2}\frac{\delta L}{\delta  \phi^{(n)}},\
    ...
\end{equation}
Then it is possible to solve from the $n$-th momentum (thanks to the degeneracy condition):
\begin{equation}
    \phi^{(n)}=F(q_1,...,q_n,\pi_n)\quad\text{such that}\quad \pi_n=\left.\frac{\delta L}{\delta \phi^{(n)}}\right|_{\phi=q_1,...,\phi^{(n-1)}=q_n,\phi^{(n)}=F}\,,
\end{equation}
and the resulting Hamiltonian,
\begin{align}
 H[q_1,...,q_n,\pi_1,...\pi_n] &= \int \Big[q_2 \pi_1 + q_3 \pi_2+...+q_n \pi_{n-1} +  F(q_1,...,q_n,\pi_n)\pi_n \Big] \dd^3x\nonumber\\
 &\qquad- L[q_1,...,q_n,F(q_1,...,q_n,\pi_n)]  \,.
\end{align}
is linear in $\{\pi_2,...,\pi_n\}$ (all momenta except $\pi_1$). These correspond to extra degrees of freedom of ghostly nature.

One can check that the Hamilton equation for $\pi_1$ is the one containing the Euler-Lagrange equation of the theory. The equations of the other momenta $\pi_i$ give the definition of $\pi_{i-1}$, the equation of $q_n$ gives the definition of $F$, and the equations of the remaining positions $q_i$ will give simply $\dot{q}_i=q_{i+1}$ \cite{Woodard2015}. See that this clearly generalizes what we obtained in \eqref{eq:Heom1}-\eqref{eq:Heom3}.

\subsection{Explicit kinetic terms for the ghosts via auxiliary fields}

In some cases, it is possible to transform an ``Ostrogradski'' ghost (non-degenerate contributions to the Lagrangian with higher-order derivatives) into an ``ordinary ghost'' (some field with negative kinetic term with no higher-order derivatives). The technique is based on a very important concept in field theory (with many applications!): auxiliary fields.

\begin{boxmathdark}
\noindent{\bf Definition. Auxiliary field}\\
We call \emph{auxiliary field} a field whose equation of motion is algebraic and admit a unique solution.
\end{boxmathdark}
When a field is auxiliary, the solution of its equation of motion can be substituted back in the action without altering the dynamics. The Lagrangians containing auxiliary fields are quadratic polynomials in them, so that the corresponding equation of motion is just a first-degree equation with a unique solution.\footnote{This definition is often relaxed to include any field with algebraic equations of motion, even if they have more than one solution.}

Let us revisit the previous example from a different perspective:

\begin{boxmathclear}

\noindent{\bf Example. An equivalent Lagrangian.}

First, we write down again the Lagrangian we used in the previous example,
\begin{equation}
  \mathcal{L}_{\phi}= {\frac 12} \partial_\mu\phi\partial^\mu\phi + {\frac \lambda 2}(\square\phi)^2 - V(\phi),\qquad \lambda \neq0.\label{eq:exampleL}
\end{equation}
Consider now the following Lagrangian density depending on two fields $\phi$ and $\chi$
\begin{equation}\label{eq:LagExampleOstr}
  \mathcal{L}_{\phi\chi}= {\frac 12} \partial_\mu\phi \partial^\mu\phi + \chi \square\phi - \frac{1}{2\lambda}\chi^2 - V(\phi)\,.
\end{equation}
We are going to show that the latter is equivalent to the former one.

On way of seeing this is by computing the set of equations of motion for both theories and check that they agree. But there is another way, which is based on the fact that the field $\chi$ is auxiliary. Its equation of motion can be solved as
\begin{equation}
  0=\frac{\partial \mathcal{L}_{\phi\chi}}{\partial \chi}= \square\phi - \frac{1}{\lambda} \chi \qquad\Rightarrow \qquad \chi = \lambda \square\phi\,,
\end{equation}
and this can be plugged back in the Lagrangian. This leads exactly to \eqref{eq:exampleL}, finishing the proof.
\end{boxmathclear}

In this example we have seen that in the reformulation of the theory in terms of $\{\phi, \chi\}$ the second-order derivatives do not appear. So, where is the ghost? The healthy degree of freedom and the extra ghost have been encoded in the pair $\{\phi, \chi\}$. Indeed, in order to see the ghost explicitly, one just need to perform an appropriate field redefinition (see also Appendix D of \cite{JoyceJain2015}):

\begin{boxmathclear}
\noindent{\bf Example. Kinetic term for the Ostrogradski ghost.}

Consider again
\begin{equation}
  \mathcal{L}_{\phi\chi}= {\frac 12} \partial_\mu\phi \partial^\mu\phi + \chi \square\phi - \frac{1}{2\lambda}\chi^2 - V(\phi)\,.
\end{equation}
After extracting a boundary term we get the following kinetic matrix:
\begin{equation}
  \mathcal{L}_{\phi\chi}= {\frac 12} (\partial_\mu\phi, \partial_\mu \chi) 
  \begin{pmatrix}1 & -1\\-1 & 0\end{pmatrix}
  \begin{pmatrix}\partial^{\mu}\phi\\\partial^{\mu}\chi\end{pmatrix}
   - \frac{1}{2\lambda}\chi^2 - V(\phi)+\partial_\mu(\chi \partial^\mu \phi)\,.
\end{equation}
The determinant of the kinetic matrix is negative, so there is a ghost in the theory. Indeed, we can diagonalize it via a field redefinition $(\phi \to \beta - \alpha,\ \chi \to \beta )$ and the Lagrangian becomes
\begin{equation}
  \mathcal{L}_{\alpha\beta}= {\frac 12} \partial_\mu\alpha \partial^\mu\alpha - {\frac 12} \partial_\mu\beta \partial^\mu\beta -  \frac{1}{2\lambda}\beta^2 - V(\beta-\alpha)\,.
\end{equation}
We can clearly see that the kinetic term of $\beta$ has the wrong sign. This field is responsible for the Hamiltonian not to be bounded from below (it can identified with the Ostrogradski ghost of the original theory).
\end{boxmathclear}

\subsection{The massive spin-1}\label{sec:spin1}

In order to describe the 3 degrees of freedom of a massive spin-1, the simplest Lorentz tensor that one can consider is a vector field $A_\mu$. The most general kinetic term for it (with Lorentz invariance) is a linear combination of the objects
\begin{equation}
  I_1:=\partial_{\mu}A_{\nu}\partial^{\mu}A^{\nu}\,,\qquad I_2:=\partial_{\mu}A_{\nu}\partial^{\nu}A^{\mu}\,,\qquad I_3:=(\partial_{\mu}A^{\mu})^{2}\,.
\end{equation}
The $I_2$ and $I_3$ are linearly dependent up to boundary terms, so we can drop one of them (the $I_2$, for convenience). Moreover, we can consider the combination
\begin{equation}
    F_{\mu\nu}(A)F^{\mu\nu}(A) = 2(\partial_{\mu}A_{\nu}\partial^{\mu}A^{\nu} - \partial_{\mu}A_{\nu}\partial^{\nu}A^{\mu} ) = 2(I_1-I_2)
\end{equation}
instead of $I_1$. Here we have introduced the antisymmetric combination $F_{\mu\nu}(A):=\partial_\mu A_\nu-\partial_\nu A_\mu$. The Lagrangian is then
\begin{equation}
\mathcal{L}_A=-\frac{1}{4}aF_{\mu\nu}(A)F^{\mu\nu}(A)-\frac{1}{2}b(\partial_{\mu}A^{\mu})^{2}+\frac{1}{2}m^{2}A_{\mu}A^{\mu}\,,\label{eq:LA}
\end{equation}
where we have introduced a mass term for the field ($m^2>0$), and two real parameters $a,b\in\mathbb{R}$. 

This theory propagates a ghostly degree of freedom due to second term, and such a ghost can be transformed into an Ostrogradski one. To do that we perform a splitting of the vector into longitudinal ($\phi$) and transversal ($B_\mu$) as follows\footnote{
   This decomposition can always be done. See \cite[sec. 8.7.1]{Schwartz2014}.}
\begin{equation}
A_{\mu}=\partial_{\mu}\phi+B_{\mu}\qquad\text{with}\qquad\partial_{\mu}B^{\mu}=0\,,
\end{equation}
which implies $\partial_{\mu}A^{\mu}=\square\phi$ and $F_{\mu\nu}(A)=F_{\mu\nu}(B)$. After this redefinition, the Lagrangian density of the theory has the form
\begin{align}
\mathcal{L}_{B\phi}&=-\frac{1}{4}aF_{\mu\nu}(B)F^{\mu\nu}(B)-\frac{1}{2}b(\square\phi)^{2}+\frac{1}{2}m^{2}(\partial_{\mu}\phi+B_{\mu})(\partial^{\mu}\phi+B^{\mu})\label{eq:LBf}\\
&=\frac{1}{2}a \delta^{ij}\big(\dot{B}_i\dot{B}_j + \partial_i B_0 \partial_j B_0 - 2 \dot{B}_i \partial_j B_0\big)-a \delta^{k[i}\delta^{j]l}\partial_{i}B_{j}\partial_{k}B_{l} \nonumber\\
&\quad-\frac{1}{2}b\Big[\ddot{\phi}^2+(\Delta \phi)^2-2 \ddot{\phi}\Delta \phi\Big]\nonumber\\
&\quad+\frac{1}{2}m^{2}(\dot{\phi}^2 - \delta^{ij}\partial_i\phi \partial_j\phi + (B_{0})^2- \delta^{ij} B_{i}B_{j})+\underbrace{\partial_\mu(m^2B^\mu\phi)}_\text{boundary term}.
\end{align}

We require $a$ to be non-trivial, because if not then we lose the dynamics for the vector part and only a scalar remains. Notice that $B_0$ does not propagate and that for $a\neq 0$ and $b\neq 0$ the Lagrangian functional is of the form $L_{B\phi}[B_0, B_i, \dot{B}_i, \phi, \dot\phi, \ddot\phi]$. It is actually enough to check the following determinant to realize that the theory has an Ostrogradski ghost:
\setlength{\arraycolsep}{2pt}
\begin{equation}\label{eq:checkdegLBfi}
  \renewcommand\arraystretch{2}
  \det\begin{pmatrix}
   \dfrac{\partial^2 \mathcal{L}_{B\phi}}{\partial\dot{B}_{i}\partial\dot{B}_{j}}&
   \dfrac{\partial^2 \mathcal{L}_{B\phi}}{\partial\dot{B}_{i}\partial\ddot{\phi}}  \\
   \dfrac{\partial^2 \mathcal{L}_{B\phi}}{\partial\ddot{\phi}\partial\dot{B}_{j}}  & 
   \dfrac{\partial^2 \mathcal{L}_{B\phi}}{\partial\ddot{\phi}\partial\ddot{\phi}}
   \end{pmatrix}
   \renewcommand\arraystretch{1.5}
   =\det\begin{pmatrix}a\mathds{1}_{3} & 0_{1\times 3} \\ 0_{3\times 1} & -b \end{pmatrix}
  \renewcommand\arraystretch{1}
   =-a^3 b \,.
\end{equation}
\setlength{\arraycolsep}{6pt}

The only way to avoid the Ostrogradski ghost (and maintain the vector degrees of freedom of $B_i$) is by requiring $b=0$. Notice that $a$ should also be positive in order to avoid a wrong sign in the kinetic term of $B_i$. So, from now on, we take $b=0$ and $a>0$. Moreover, we can redefine the fields and the mass parameter absorbing this $a$ in order to get the canonical normalization for the kinetic term of the vector part, 
so that \eqref{eq:LA} becomes:
\begin{equation}
\mathcal{L}_\text{Proca}=-\frac{1}{4}F_{\mu\nu}(A)F^{\mu\nu}(A)+\frac{1}{2}m^{2}A_{\mu}A^{\mu}\,.
\end{equation}
This is the \emph{Proca Lagrangian} and propagates the three degrees of freedom needed for the quantum field theory of a massive spin-1 particle: two in the transversal part of the field $B_\mu$ and one in the longitudinal part $\phi$. Observe that the dynamics of $\phi$ is controlled by the mass term. In the massless case (Maxwell) such a degree of freedom is absent.

It is also interesting to notice that the Proca Lagrangian is completely equivalent to the so called  \emph{St\"uckelberg Lagrangian} (after the introduction of a new field $A_\mu \to A_\mu+\partial_\mu \varphi$):
\begin{equation}
\label{eq:ProcaActionStueckelberg}
\mathcal{L}_\text{St\"uck}=-\frac{1}{4}F_{\mu\nu}(A)F^{\mu\nu}(A)+\frac{1}{2}m^{2}(\partial_{\mu}\varphi+A_{\mu})(\partial^{\mu}\varphi+A^{\mu})\,.
\end{equation}
This Lagrangian has a local symmetry
\begin{equation}
\label{eq:StueckelbergGauge}
A_{\mu}\to A_{\mu}+\partial_{\mu}g ~\,,\qquad\varphi\to\varphi-g\,.
\end{equation}
so any given $\varphi$ is redundant since it can be totally absorbed by choosing an appropriate $g$. If we fix the gauge in which $\varphi=0$, we recover the Proca Lagrangian, so $\mathcal{L}_\text{St\"uck}$ also describes the 3 degrees of freedom of a massive spin-1. In the St\"ucekelberg case, the gauge symmetry is telling us that the longitudinal mode that we can extract from $A_\mu$ does not introduce ghosts, because it can be absorbed in $\varphi$, which has a healthy kinetic term. The Proca field then has 1 longitudinal degree of freedom (thanks to the non-vanishing mass) and 2 transversal degrees of freedom. 

It is interesting to notice how the gauge symmetries are connected with the absence of Ostrogradski ghosts coming from the longitudinal part of the fields. In the massless case, the $\mathrm{U}(1)$ gauge symmetry $A_{\mu}\to A_{\mu}+\partial_{\mu}g$ prevents the Maxwell Lagrangian from being pathological. The same thing happens in the Fierz-Pauli Lagrangian for a massless spin-2 field, where the ghosts are absent and the kinetic sector corresponds to the one with the gauge symmetry $h_{\mu\nu}\to h_{\mu\nu}+\partial_{\mu}\xi_{\nu}+\partial_{\nu}\xi_{\mu}$. Observe that, in these two cases, what the gauge freedom is telling us is that the longitudinal parts (the ones connected to the Ostrogradski instability) are pure gauge, i.e. redundant/unphysical. In other words, they do not appear in the Lagrangian, proving that the theories are safe from these problems.

\newpage
\begin{center}
  \textbf{LIST OF PROBLEMS FOR LECTURE 2}    
\end{center}
\addcontentsline{toc}{section}{List of problems for Lecture 2}

\vspace{.5cm}

{\small \noindent {\it Conventions}: we choose the signature $\eta_{\mu\nu} = {\rm diag}(+1,-1,-1,-1)$.}

\vspace{.3cm}

\noindent{}\textbf{Problem 1.} \textit{Ostrogradski procedure}
\vspace{.3cm}

\noindent{}The following Lagrangian density describes a field theory with two fields $\varphi$ and $\psi$, 
    \begin{equation}
       \mathcal{L}= \frac{1}{2}(a\partial_\mu \partial_\nu\varphi\partial^\mu\partial^\nu \varphi + \partial_\mu \psi \partial ^\mu \psi + b \partial_\mu \psi \partial^\mu\varphi + b \psi \square \varphi)\ \dd^3 x\,,
    \end{equation}
    where $\square:= \eta^{\mu\nu}\partial_\mu \partial_\nu$.
\begin{enumerate}[label=\alph*)]
    \item Is it possible to simplify the Lagrangian?
    \item The Hessian matrix is given by:
    \begin{equation}
        \begin{pmatrix}\dfrac{\partial^{2}\mathcal{L}}{\partial\ddot{\varphi}^{2}} & \dfrac{\partial^{2}\mathcal{L}}{\partial\dot{\psi}\partial\ddot{\varphi}}\\
\dfrac{\partial^{2}\mathcal{L}}{\partial\ddot{\varphi}\partial\dot{\psi}} & \dfrac{\partial^{2}\mathcal{L}}{\partial\dot{\psi}^{2}}
\end{pmatrix}
    \end{equation}
    Calculate it and show that whenever $a\neq 0$, the theory propagates more than 2 degrees of freedom.
    \item Under the condition $b=0$, compute the Ostrogradski Hamiltonian. Is the Ostrogradski theorem fulfilled in this case? Why?
\end{enumerate}

\vspace{.6cm}

\noindent{}\textbf{Problem 2.} \textit{Field redefinitions and detection of ghosts}
\vspace{.3cm}

\noindent{}Consider a theory depending on $\{\Psi^A, A_\mu, B_\nu\}$, where $\{\Psi^A\}$ is a family of arbitrary fields (vectors, scalars, tensors...) that propagate healthily as dictated by the Lagrangian
    \begin{equation}
        \mathcal{L}= a\ F_{\mu\nu}(A)F^{\mu\nu}(B) + (\text{kinetic terms for }\,\Psi^A) + V(\Psi^A, A_\mu, B_\nu)
    \end{equation}
where $F_{\mu\nu}(X):=\partial_\mu X_\nu- \partial_\nu X_\mu$, $a$ is a real number and $V$ does not depend on the derivatives of the fields.
\begin{enumerate}[label=\alph*)]
    \item Diagonalize the kinetic sector of $A_\mu$ and $B_\mu$ by doing the field redefinition,
    \begin{equation}
        A_\mu = U_\mu + V_\mu\,\qquad  B_\mu = U_\mu - V_\mu\,,
    \end{equation}
    and show that the presence of the first term (i.e., $a\neq 0$) automatically indicates that one of the vectors is a ghost.
    
    {\it Hint}: Remember that the canonical normalization for a vector is $-\frac{1}{4}F_{\mu\nu}F^{\mu\nu}$.
    
    \item Indeed it is not necessary to perform the previous field redefinition. The fact that there is a ghost can be directly seen from the kinetic matrix (even if it has not been diagonalized). To see this, rewrite the kinetic sector of $A_\mu$ and $B_\mu$ as
    \begin{equation}
        a\ F_{\mu\nu}(A)F^{\mu\nu}(B)  = -\frac{1}{4} F_{\mu\nu}(X^i)\ M_{ij}\ F^{\mu\nu}(X^j)
    \end{equation}
    with $X^i=(A,B)$, and calculate the determinant of the kinetic matrix $M_{ij}$. What can we conclude from it?
    
    \item Assume that there are three vectors (instead of two) whose kinetic sector is separated from the rest of the theory. What can be said about the presence of ghosts depending on the sign of the determinant of the kinetic matrix? 
\end{enumerate}

\vspace{.6cm}

\noindent{}\textbf{Problem 3.} \textit{Dangerous interaction terms between vectors}
\vspace{.3cm}

\noindent{}Consider a theory with two interacting massive vector fields $\{A_\mu,B_\mu\}$ with (a priori) safe kinetic sector:
\begin{equation}
    \mathcal{L} = -\frac{1}{4}F_{\mu\nu}(A)F^{\mu\nu}(A) -\frac{1}{4}F_{\mu\nu}(B)F^{\mu\nu}(B) + \frac{1}{2}m^2_AA_\mu A^\mu + \frac{1}{2}m_B^2B_\mu B^\mu + \mathcal{L}_\text{int}\,.
\end{equation}
Show that a self-interaction of the type
    \begin{equation}
        \mathcal{L}_\text{int}^{(1)}= A_\mu A^\mu \partial_\nu A^\nu\,,
    \end{equation}
    for one (or both) vectors is a healthy interaction, whereas the interactions that mix the vectors,
    \begin{equation}
        \mathcal{L}_\text{int}^{(2)}= A_\mu A^\mu \partial_\nu B^\nu\qquad\text{and}\qquad \mathcal{L}_\text{int}^{(3)}= A_\mu B^\mu \partial_\nu A^\nu,
    \end{equation}
    introduce a ghost. \\
    {\it Hint:} First extract the longitudinal part of each of the vectors, i.e., perform $X_\mu\to \partial_\mu \phi + Y_\mu$, where the parts $Y_\mu$ are transversal ($\partial_\mu Y^\mu=0$), and analyze the higher-order terms (in time derivatives) for the longitudinal parts $\phi$'s. Then, study the contribution of these longitudinal parts to the Hessian.

\vspace{.6cm}

\noindent{}\textbf{Problem 4.} \textit{Theory with two scalars and second-order derivatives.}
\vspace{.3cm}

\noindent{}Consider the Lagrangian
\begin{equation}
    \mathcal{L} = \frac{1}{2}\ddot{\phi}^2+\frac{1}{2}\ddot{\psi}^2+\ddot{\phi}\ddot{\psi}- \frac{1}{2}m^2 (\phi^2+\psi^2)\qquad m^2>0\,.
\end{equation}
How many degrees of freedom does this theory propagate? Is it ghost-free?\\
{\it Hint:} Check first the Hessian, and then try to find a field redefinition that simplifies the higher-order part.

\newpage

\lecture{Examples of instabilities in gravity theories}
\addcontentsline{toc}{section}{Examples of instabilities in gravity theories}

\subsection{Stability issues in modified theories of gravity}

In this lecture we shall use the knowledge of the previous chapters to analyze the possible appearance of ghosts in modifications of GR.

A Hamiltonian analysis of the Einstein-Hilbert action reveals that the theory is healthy at the full non-linear level \cite{poisson2004relativist}. Nevertheless, when modifying such an action, we may expect some pathological behavior. Let us explain the reasoning behind this in the following.

The Lovelock theorem states that: ``The only possible equations of motion from a local gravitational action which contains only second derivatives of the four-dimensional spacetime metric are the Einstein field equations''. From such a statement we can derive a straightforward conclusion: if we make a local, four-dimensional modification of GR, we are introducing higher derivatives and/or extra fields (scalar, vector and tensors). On the one hand, from the previous lecture we know that having higher than second derivatives in the action almost certainly signals a pathological behavior. On the other hand, the extra fields present in the action could be ghosts or tachyons depending on the parameters of the theory. 

Therefore, when exploring possible modifications of GR, instabilities of different kinds are expected. Here we present some examples of modified theories of gravity and their known pathological/safe behavior. We follow the discussion in Sec. 7.2 of \cite{JCAthesis}:
\begin{itemize}
\item First we discuss about metric theories with higher-order terms in the curvature. It is known that only Lovelock theories have generically the same spectrum as General Relativity, i.e., a massless graviton. Any other higher-order curvature theory propagates additional typically problematic degrees of freedom (generically, a massive ghostly spin-2  and a scalar). An example of this is (Cosmological) Einsteinian Cubic gravity, as we will see in Section \ref{sec:SC}, which contains a ghost that becomes strongly coupled around spatially-flat cosmological backgrounds \cite{BeltranJCA2021}. \\
However, a miracle happens in some degenerate cases, and these generic problems are absent. A good example of this is the $f(\mathring{R})$ family of theories. This can be seen directly in the Einstein frame (the result is just GR plus a scalar field) or by taking a Lagrangian of the type $\mathring{R}^2+a \mathring{R}_{\mu\nu} \mathring{R}^{\mu\nu}$ and analyzing the limit $a\rightarrow0$. It can be seen that the mass of the extra spin-2 goes to infinity without rendering any pathologies \cite{BeltranJCA2021}. This result for $\mathring{R}^2$ can be extended to the entire family of $f(\mathring{R})$ \cite{Woodard2007}.\\
Another miraculous case is a Lagrangian consisting on an arbitrary function of the Gauss-Bonnet invariant, called $f(G)$-gravity (see e.g. \cite{NojiriOdintsov2005, CognolaElizalde2006,DeFeliceTsujikawa2009}). Here again, only the massless graviton and a safe scalar propagate \cite{Kobayashi2011}.

\item Ricci-based gravity theories with projective symmetry (see Section \ref{sec:RB}) are known to be equivalent GR up to a field redefinition (i.e., they admit an Einstein frame), so only the graviton propagates. However, when the action also includes the antisymmetric part of the Ricci tensor, the broken projective symmetry leads to new ghostly degrees of freedom \cite{BeltranDelhom2019,BeltranDelhom2020}. We will discuss this in more detail in Section \ref{sec:RB}.

\item The non-linear extensions of the teleparallel equivalents, $f(T)$ \cite{GolovnevKoivisto2018,BengocheaFerraro2008, LiSotiriouBarrow2010} and $f(Q)$ \cite{BeltranHeisenbergKoivisto2018a, BeltranHeisenbergPekar2019} also suffer from instabilities. The key to understand the origin of these new degrees of freedom is the following: the teleparallel equivalents of GR, $T$ and $Q$ exhibit some special symmetries that lead to a propagating  graviton. To be precise, such groups of transformations only leave the action invariant up to a boundary term. Consequently, these boundary terms cannot escape from the function $f$ in the non-linear extension. The symmetries are then lost, and new degrees of freedom appear. \\
For instance, it has be shown that maximally symmetric backgrounds are strongly coupled in $f(Q)$ \cite{BeltranHeisenbergPekar2019}. For $f(T)$, the situation is even worse, since the strong coupling problem affects more general cosmological backgrounds \cite{GolovnevKoivisto2018}.

\item The quadratic Poincar\'e Gauge (PG) gravity Lagrangian, which will be explored in Section \ref{sec:PG}, contains ghosts and tachyons for generic values of the parameters (see e.g. \cite{BlagojevicCvetkovic2018}). In fact, in \cite{Yo:1999ex,Yo:2001sy} (and more recently in \cite{BeltranJimenez:2019hrm}), authors found that the only modes that could propagate safely were the two spin-0 with different parity. Similar problems are expected in quadratic metric-affine gravity (see for example the recent work \cite{JCMT2022}).
\end{itemize}

All of the examples above are summarized in Table \ref{tab:safe} for the healthy theories, and in Table \ref{tab:pathological} for the unstable ones.

\renewcommand\arraystretch{1.2}

\begin{table}
    \centering
    \begin{tabular}{|c|c|}
\hline 
Theory & Field content \tabularnewline
\hline 
\hline 
Lovelock (GR, GB,...) & Graviton\tabularnewline
\hline 
$f(\mathring{R})$, $f(\text{GB})$ & Graviton + scalar\tabularnewline
\hline 
$f(R)$ & Graviton {[}+ non-dynamical scalar{]}\tabularnewline
\hline 
Horndeski gravity & Graviton + scalar\tabularnewline
\hline 
Generalized Proca & Graviton + (massive) vector\tabularnewline
\hline 
Ricci-Based Gravity & Graviton + scalar\tabularnewline
\hline 
\end{tabular}
    \caption{Some examples of modified theories of gravity which do not introduce instabilities under some restrictions.}
    \label{tab:safe}
\end{table}

\begin{table}
    \centering
    \begin{tabular}{|c|c|}
\hline 
Theory & Some known pathologies\tabularnewline
\hline 
\hline
Massive gravity (original formulation) & Boulware-Deser ghost \tabularnewline
\hline 
Ricci-Based Gravity with $R_{[\mu\nu]}$ & Ghost (projective mode)\tabularnewline
\hline 
Generic higher curvature gravity  & Potential ghosts (massive spin-2, scalar) \tabularnewline
\hline 
$f(T)$ & Strong coupling problem in FLRW\tabularnewline
\hline 
$f(Q)$ & Strong coupling problem in Max. Sym.\tabularnewline
\hline 
Quadratic PG gravity & Ghosts and tachyons (and strong c.) \tabularnewline
\hline 
General quadratic MAG gravity & Similar problems as in PG\tabularnewline
\hline 
\end{tabular}
    \caption{Paradigmatic examples of modifications of GR and their known pathological behavior.}
    \label{tab:pathological}
\end{table}

\renewcommand\arraystretch{1}

In the following sections we shall study in detail the appearance of instabilities in PG gravity \cite{BeltranJimenez:2019hrm}, (Cosmological) Einsteinian Cubic gravity \cite{BeltranJCA2021}, and Ricci-based theories \cite{BeltranDelhom2019,BeltranDelhom2020}.

\subsection{Stable restrictions of quadratic Poincaré Gauge gravity}
\label{sec:PG}

With the knowledge acquired in the previous lectures we are ready to study the stability of PG gravity. We shall summarize the results of \cite{BeltranJimenez:2019hrm}, where the stable modes of propagation of the action quadratic PG action were found. We shall unveil the presence of pathological terms in a background-independent approach just by looking at the interactions of the different torsion components. 

To start with, let us present the action of quadratic PG gravity, which is given by
\begin{align}
\label{actionPGT}
S_{{\rm PG}}=&\int {\rm d}^{4}x\sqrt{-g}\left(a_{0}R+a_{1}T_{\mu\nu\rho}T^{\mu\nu\rho}+a_{2}T_{\mu\nu\rho}T^{\nu\rho\mu}+a_{3}T_{\mu}T^{\mu}+b_{1}R^{2}\right.
\nonumber\\
&\qquad+b_{2}R_{\mu\nu\rho\sigma}R^{\mu\nu\rho\sigma}+b_{3}R_{\mu\nu\rho\sigma}R^{\rho\sigma\mu\nu}+b_{4}R_{\mu\nu\rho\sigma}R^{\mu\rho\nu\sigma}
\nonumber\\
&\qquad+\left.b_{5}R_{\mu\nu}R^{\mu\nu}+b_{6}R_{\mu\nu}R^{\nu\mu}\right),
\end{align}
where the $a_i$ and $b_i$ are constant parameters of the theory.

This theory has clearly more degrees of freedom than GR, due to the quadratic curvature and torsion terms. To analyse those introduced by the torsion tensor, it is customary to decompose the torsion in three terms
\begin{equation}
\label{decomposition}
\begin{cases}
\textrm{Trace vector:}\,\, T_{\mu}:=T^{\nu}{}_{\mu\nu},\\
\,\\
\textrm{Axial vector:}\,\, S_{\mu}:=\varepsilon_{\mu\nu\rho\sigma}T^{\nu\rho\sigma},\\
\,\\
\textrm{Tensor}\,\, ^{\rho}{}_{\mu\nu},\,\,{\rm such\,that} \;\;q^{\nu}{}_{\mu\nu}=0\,\, \textrm{and}\,\, \varepsilon_{\mu\nu\rho\sigma}q^{\nu\rho\sigma}=0,
\end{cases}
\end{equation}
such that
\begin{equation}
\label{decomposition2}
T^{\rho}{}_{\mu\nu}=\frac{1}{3}\left(T_{\mu}\delta_{\nu}^{\rho}-T_{\nu}\delta_{\mu}^{\rho}\right)+\frac{1}{6}\varepsilon^{\rho}{}_{\mu\nu\sigma}S^{\sigma}+q^{\rho}{}_{\mu\nu}.
\end{equation}
This decomposition turns out to be very useful, thanks to the fact that the three terms in \eqref{decomposition} propagate different dynamical off-shell degrees of freedom.

Let us also remember that every non-symmetric and metric compatible connection, it can be related with the Levi-Civita connection as
\begin{equation}
\label{LCTorsion}
\Gamma^{\alpha}{}_{\mu\nu}=\mathring{\Gamma}^{\alpha}{}_{\mu\nu}+K^{\alpha}{}_{\mu\nu},
\end{equation}
where $\mathring{\Gamma}$ is the Levi-Civita connection and 
\begin{equation}
\label{contorsion}
K^\alpha{}_{\mu\nu}:=\frac12 \Big(T^\alpha{}_{\mu\nu}+T_{\mu}{}^\alpha{}_\nu+T_\nu{}^\alpha{}_\mu\Big)
\end{equation}
is the {\emph{contortion tensor}}. The decomposition of the quantities related to the total connection using \eqref{LCTorsion} is known as a \emph{post-Riemannian expansion}.\\

With respect to the stability of the Lagrangian \eqref{actionPGT}, in order to avoid ghosts already for the graviton when the torsion is set to zero, we will impose the recovery of the Gauss-Bonnet term in the limit of vanishing torsion. In 4 dimensions we can use the topological nature of the Gauss-Bonnet term to remove one of the parameters. More explicitly, we have
\begin{eqnarray}
\label{eq:PGactionzeroT}
\mathcal{L}_{\rm PG}\big\vert_{T=0}=a_{0}\mathring{R}+\left(b_{2}+b_{3}+\frac{b_{4}}{2}\right)\mathring{R}_{\mu\nu\rho\sigma}\mathring{R}^{\mu\nu\rho\sigma}+\left(b_{5}+b_{6}\right)\mathring{R}_{\mu\nu}\mathring{R}^{\mu\nu}+b_{1}\mathring{R}^{2},
\end{eqnarray}
so the Gauss-Bonnet term for the quadratic sector is recovered upon requiring
\begin{equation}
\label{con1}
b_{5}=-4b_{1}-b_{6},\quad b_{4}=2(b_{1}-b_{2}-b_{3}).
\end{equation}

In the following subsection we shall show that imposing stability in the torsion vector modes reduces drastically the parameter space of PG gravity. 

\subsubsection{Ghosts in the vector sector}

In 4 dimensions, a vector field $A_{\mu}$ has four components: one temporal $A_{0}$, and three spatial $A_{i}$, with $i=1,2,3$. However, as we saw in the previous lecture, they cannot propagate at the same time without introducing a ghost d.o.f.. In particular for any theory describing a massive vector, like the ones present in the PG action, we must require the following conditions in order to avoid ghosts \cite{Heisenberg2014,Heisenberg2018}:
\begin{itemize}
\item {\bf{The equations of motion must be of second order}}. As we explained in the previous lecture, this is because the Ostrogradski theorem predicts ghosts for higher-order equations of motion.

\item {\bf{The temporal component of the vector field $A_0$ should not be dynamical}}. Therefore, the massive vector under this ghost-free condition would only propagate three degrees of freedom, which is exactly the ones that the massive spin-1 representation of the Lorentz group can propagate. 
\end{itemize}
Following these prescriptions, in this subsection we shall constrain the parameter space of PG gravity by imposing stability in the two massive spin-1 fields that are part of the particle spectrum of this theory.\\
In order to do so, we look at the vector sector containing the trace $T_\mu$ and the axial component $S_\mu$ of the torsion, while ignoring the pure tensor part $q^{\rho}{}_{\mu\nu}$ for the moment. Plugging the decompositions \eqref{LCTorsion} and \eqref{decomposition2} into the quadratic PG Lagrangian \eqref{actionPGT} we obtain
\begin{eqnarray}
\Lag_{\rm v}&=&-\frac29\big(\kappa-\beta\big)\mT_{\mu\nu}\mT^{\mu\nu}+\frac{1}{72}\big(\kappa-2\beta\big)\mS_{\mu\nu}\mS^{\mu\nu}+\frac12m_T^2T^2+\frac12m_S^2S^2+\frac{\beta}{81} S^2 T^2\nonumber\\
&&+\frac{4\beta-9b_2}{81}\Big[(S_\mu T^\mu)^2+3S^\mu S^\nu \nablab_\mu T_\nu\Big]+\frac{\beta}{54} S^2\nablab_\mu T^\mu+\frac{\beta-3b_2}{9}S^\mu T^\nu\nablab_\mu S_\nu \nonumber\\
&&+\frac{\beta-3b_2}{12} (\nablab_\mu S^\mu)^2+\frac{\beta}{36}\Big(2\Gb^{\mu\nu}S_\mu S_\nu+\Rb S^2\Big),
\label{eq:PGaction2}
\end{eqnarray}
where $\mT_{\mu\nu}:=2\partial_{[\mu} T_{\nu]}$ and $\mS_{\mu\nu}:=2\partial_{[\mu} S_{\nu]}$ are the fieldstrengths of the trace and axial vectors respectively and we have defined
\begin{eqnarray}
\kappa&:=&4b_1+b_6\, ,\\
\beta&:=&b_1+b_2-b_3\, ,\\
m_T^2&:=&-\frac23\big(2a_0-2a_1+a_2-3a_3\big)\, ,\\
m_S^2&:=&\frac{1}{12}\big(a_0-4a_1-4a_2\big).
\end{eqnarray}
In order to arrive at the final expression \eqref{eq:PGaction2} we have used the Bianchi identities to eliminate terms containing $\Rb_{\mu\nu\rho\sigma}\epsilon^{\alpha\nu\rho\sigma}$ and express $\Rb_{\mu\nu\rho\sigma}\Rb^{\mu\rho\nu\sigma}=\frac12\Rb_{\mu\nu\rho\sigma}\Rb^{\mu\nu\rho\sigma}$. We have also dropped the Gauss-Bonnet invariant of the Levi-Civita connection and the total derivative $\varepsilon_{\mu\nu\alpha\beta} \mS^{\mu\nu}\mT^{\alpha\beta}$. Moreover, we have made a few integrations by parts and used the commutator of covariant derivatives. We can see that the parameter $b_1$ does not play any role since it simply corresponds to the irrelevant Gauss-Bonnet coupling constant.

If we look at the pure trace sector $T_\mu$ of \eqref{eq:PGaction2}, we observe that it does not contain non-minimal couplings or divergence square terms. Unlike the torsion trace, the axial component $S_\mu$ shows very worrisome terms that appear in the three following ways:
\begin{itemize}
\item  The first pathological term is $(\nablab_\mu S^\mu)^2$ that introduces a ghostly d.o.f. because it clearly makes the four components of the vector propagate. We shall get rid of it by imposing $\beta=3b_2$. 

\item The non-minimal couplings to the curvature which are not Horndeski-like are known to lead to ghostly d.o.f.'s. The presence of these instabilities shows in the metric field equations where again all the components of the vector will enter dynamically, hence revealing its problematic dynamics. An exception is the coupling to the Einstein tensor that avoids generating these second time derivatives due to its divergenceless property. For this reason we have explicitly separated the non-minimal coupling to the Einstein tensor in \eqref{eq:PGaction2}. It is therefore clear that we need to impose the additional constraint $\beta=0$ to guarantee the absence ghosts, which, in combination with the above condition $\beta=3b_2$, results in $\beta=b_2=0$.

\item Furthermore, there are other interactions in \eqref{eq:PGaction2} with a generically pathological character schematically given by $S^2\nabla T$ and $ST\nabla S$. Although these may look like safe vector Galileon-like interactions, actually the fact that they contain both sectors makes them dangerous. This can be better understood by introducing St\"uckelberg fields, so we effectively have $T_\mu\rightarrow\partial_\mu T$ and $S_\mu\rightarrow\partial_\mu S$ with $T$ and $S$ the scalar and pseudo-scalar St\"uckelbergs. The interactions become of the form $(\partial S)^2\partial^2 T$ and $\partial T\partial S\partial^2T$ that, unlike the pure Galileon interactions, generically give rise to higher-order equations of motion and, therefore, Ostrogradski instabilities. Nevertheless, we can see that the avoidance of this pathological behavior does not introduce new constraints on the parameters, since the coefficients in front of them in \eqref{eq:PGaction2} are already zero if we take into account the two previous stability considerations. 

\end{itemize}

Let us note that the extra constraint $\beta=0$ genuinely originates from the quadratic curvature interactions in the PGT Lagrangian. Such interactions induce the non-minimal couplings between the axial sector and the graviton, as well as the problematic non-gauge-invariant derivative interactions. Also, {\bf{this constraint cannot be obtained from a perturbative analysis on a Minkowski background}} because, in that case, these interactions will only enter at cubic and higher orders so that the linear analysis is completely oblivious to it.

We can see that the stability conditions not only remove the obvious pathological interactions mentioned before, but they actually eliminate all the interactions and only leave the free quadratic part
\begin{equation}
\Lag_{\rm v}\big\vert_{b_2,\beta=0}=-\frac{2}{9}\kappa\mT_{\mu\nu}\mT^{\mu\nu}+\frac12m^2_T T^2+\frac{1}{72}\kappa\mS_{\mu\nu}\mS^{\mu\nu}+\frac12m^2_S S^2
\label{eq:PGaction4}
\end{equation}
where we see that the kinetic terms for $T_\mu$ and $S_\mu$  have the same normalization but with opposite signs, hence leading to the unavoidable presence of a ghost. Therefore, the only stable possibility is to exactly cancel both kinetic terms. Consequently, the entire vector sector becomes non-dynamical.\\

Now that we have shown that the vector sector must trivialize in stable PGTs, we can return to the full torsion scenario by including the pure tensor sector $q^\rho{}_{\mu\nu}$. Instead of using the general decomposition \eqref{decomposition2}, it is more convenient to work with the torsion directly for our purpose here. We can perform the post-Riemannian decomposition for the theories with a stable vector sector to obtain
\begin{equation}
\label{Stablenonprop}
\mathcal{L}_{\rm stable}=a_{0}\mathring{R}+b_{1}\GB+a_{1}T_{\mu\nu\rho}T^{\mu\nu\rho}+a_{2}T_{\mu\nu\rho}T^{\nu\rho\mu}+a_{3}T_\mu T^\mu.
\end{equation}
The first term is just the usual Einstein-Hilbert Lagrangian, modulated by $a_0$, while the second term corresponds to the topological Gauss-Bonnet invariant for a connection with torsion, so we can safely drop it in four dimensions and, consequently, the first two terms in the above expression simply describe GR. The rest of the expression clearly shows the non-dynamical nature of the full torsion so that having a stable vector sector also eliminates the dynamics for the tensor component, therefore making the full connection an auxiliary field.

\subsubsection{Constructing a stable Poincar\'e Gauge theory}

The precedent subsection has been devoted to showing the presence of ghosts in general quadratic PG, when at least one of the spin-1 fields in $T_\mu$ or $S_\mu$ is propagating. Although this is a drawback for a very general class of theories, we can construct safe theories if we make these spin-1 fields non-propagating. In particular, in the following we will show a specific class of ghost-free theory which propagates the pseudo-scalar mode given by the longitudinal part (spin-0) of the axial torsion vector.

We can ask whether there is some non-trivial healthy theory described by \eqref{eq:PGaction2} where the scalar is associated to the axial vector. The answer is indeed affirmative, and in order to prove such a result we simply need to impose the vanishing of the Maxwell kinetic terms that results in the following conditions
\begin{equation}
\kappa=0\quad {\text{and}}\quad \beta=0.
\end{equation}
Imposing these conditions, performing a few integrations by parts and dropping the Gauss-Bonnet term, the Lagrangian then reads
\begin{equation}
\Lag_{\rm Holst}=a_0\Rb+\frac12m_T^2 T^2+\frac12m_S^2S^2+\alpha\left[(\nablab_\mu S^\mu)^2-\frac43 S_\mu T^\mu\nablab_\nu S^\nu+\frac49(S_\mu T^\mu)^2\right], 
\label{eq:Holst1}
\end{equation}
with $\alpha:=-\frac{b_2}{4}$. 

At this moment, let us use the definition of the so-called Holst term,\footnote{
    Although this term is commonly known as the Holst term, due to the research article of Soren Holst in 1995 \cite{Holst:1995pc}, in the context of torsion gravity it was first introduced by R. Hojman {\emph{et. al.}} in 1980 \cite{Hojman:1980kv}.}
which is given by $\mathcal{H}:=\epsilon^{\mu\nu\rho\sigma} R_{\mu\nu\rho\sigma}$ and whose post-Riemannian expansion is
\begin{equation}
\mathcal{H}=\frac23 S_\mu T^\mu-\nablab_\mu S^\mu
\label{eq:HoldstPR}
\end{equation}
where we have used that $\epsilon^{\mu\nu\rho\sigma} \Rb_{\mu\nu\rho\sigma}=0$ by virtue of the Bianchi identities. Thus, it is obvious that the Lagrangian can be expressed as
\begin{equation}
\Lag_{\rm Holst}=a_0\Rb+\frac12m_T^2 T^2+\frac12m_S^2S^2
+\alpha \mathcal{H}^2 .
\label{eq:Holst2}
\end{equation}
We will understand the nature of this scalar by introducing an auxiliary field $\phi$ to rewrite \eqref{eq:Holst2} as
\begin{equation}
\Lag_{\rm Holst}=a_0\Rb+\frac12m_T^2 T^2+\frac12m_S^2S^2-\alpha\phi^2+2\alpha\phi\epsilon^{\mu\nu\rho\sigma} R_{\mu\nu\rho\sigma}.
\label{eq:Holst3}
\end{equation}
As we shall show now, the field $\phi$ is dynamical and corresponds to the pseudo-scalar mode in the PG Lagrangian. 

At this moment, we can introduce the post-Riemannian expansion \eqref{eq:HoldstPR} into the Lagrangian, obtaining
\begin{equation}
\Lag_{\rm Holst}=a_0\Rb+\frac12m_T^2 T^2+\frac12m_S^2S^2-\alpha\phi^2+2\alpha\phi\left(\frac23 S_\mu T^\mu-\nablab_\mu S^\mu\right).
\end{equation}
The correspondent equations for $S^\mu$ and $T^\mu$ are 
\begin{align}
m_S^2S_\mu+\frac{4\alpha\phi}{3}T_\mu+2\alpha\partial_\mu\phi&=0,\label{eq:S}\\
m_T^2T_\mu+\frac{4\alpha\phi}{3}S_\mu&=0,
\end{align}
respectively. For $m_T^2\neq0$,\footnote{The singular value $m_T^2=0$ leads to uninteresting theories where all the dynamics is lost so we will not consider it any further here.} we can algebraically solve these equations as
\begin{align}
T_\mu&=-\frac{4\alpha\phi}{3m_T^2}S_\mu,\\
S_\mu&=-\frac{2\alpha\partial_\mu\phi}{m_S^2-\left(\frac{4\alpha\phi}{3m_T}\right)^2},\label{eq:solS}
\end{align}
which we can plug into the Lagrangian to finally obtain
\begin{equation}
\Lag_{\rm Holst}=a_0\Rb-\frac{2\alpha^2}{m_S^2-\left(\frac{4\alpha\phi}{3m_T}\right)^2}(\partial\phi)^2-\alpha\phi^2.
\label{eq:Holst4}
\end{equation}
This equivalent formulation of the theory where all the auxiliary fields have been integrated out explicitly shows the presence of a propagating pseudo-scalar field. Moreover, we can see how including the pure tensor part $q^\rho{}_{\mu\nu}$ into the picture does not change the conclusions because it contributes to the Holst term as
\begin{equation}
\mathcal{H}=\frac23 S_\mu T^\mu-\nablab_\mu S^\mu+\frac12\epsilon_{\alpha\beta\mu\nu}q_\lambda{}^{\alpha\beta}q^{\lambda\mu\nu}.
\end{equation}
This shows that $q_{\rho\mu\nu}$ only enters as an auxiliary field whose equation of motion trivializes it.

The stability constraints on the parameters can now be obtained very easily. From \eqref{eq:Holst4} we can realize that $\alpha$ must be positive to avoid having an unbounded potential from below. On the other hand, the condition to prevent $\phi$ from being a ghost depends on the signs of $m_S^2$ and  $m_T^2$,  which are not defined by any stability condition so far. We can distinguish the following possibilities:
\begin{itemize}
\item $m_S^2>0$: We then need to have $1-\left(\frac{4\alpha\phi}{3m_Tm_S}\right)^2>0$. For $m_T^2<0$ this is always satisfied, while for $m_T^2>0$ there is an upper bound for the value of the field given by $\vert\phi\vert<\vert\frac{3m_Sm_T}{4\alpha}\vert$.

\item $m_S^2<0$: The ghost-freedom condition is now $1-\left(\frac{4\alpha\phi}{3m_Tm_S}\right)^2<0$, which can never be fulfilled if $m_T^2>0$. If $m_T^2<0$ we instead have the lower bound $\vert\phi\vert>\vert\frac{3m_Sm_T}{4\alpha}\vert$.
\end{itemize}

%%%%%%%%%%%%%%%%%%%%%%%%%%%%%%%%%%%%%%%%%%%%%%%%%%%%%%%%%%%%%%%%%%%%%%%%%%%%%%%%%%%%%%%%%%%%%%%%%%%%%%%%

\subsection{Strong coupling in cosmological Einsteinian Cubic gravity}
\label{sec:SC}

This section is devoted to a particular kind of cubic (metric) theory of gravity and the presence of instabilities due to strongly coupled degrees of freedom. To start with, we revise the definition of these theories, as well as some basic properties that are needed for our analysis.

\subsubsection{Introduction to Einsteinian Cubic Graviti(es)}

\emph{Einsteinian Cubic Gravity} \cite{BuenoCano2016b} is a higher-order curvature theory of gravity that possesses the same linear spectrum as General Relativity  around maximally symmetric spacetimes in arbitrary dimension \cite{OlivaRay2010, MyersRobinson2010, BuenoCano2016b} (i.e., only the graviton propagates). These theories were extended in \cite{ArciniegaEdelstein2018}  to the so-called \emph{Cosmological Einsteinian Cubic Gravity} (CECG), whose linear spectrum coincides also with the one of GR not only around maximally symmetric spaces but in arbitrary FLRW spacetimes (see e.g. \cite{ArciniegaEdelstein2018, CisternaGrandi2020,  ArciniegaBuenoCano2018}).

As it is well-known, Lovelock terms are the only higher-order curvature theories with the same field content as GR around arbitrary backgrounds \cite{Lovelock1970,Lovelock1971}. Consequently, the cubic theories we mentioned above must contain additional d.o.f.'s, which will be connected to Ostrogradski instabilities due to the higher-order nature of the equations. Einsteinian Cubic gravities are specific cubic polynomial of the Riemann tensor such that the field equations become second order around specific backgrounds. Therefore, less d.o.f.'s are propagating there, what is an indicator of a strong coupling problem. In this section, based on the work \cite{BeltranJCA2021}, we revise the implications of such a problem\footnote{
   These problems were already discussed for the case of inflationary solutions in \cite{PookkillathDeFelice2020}.} 
around spatially flat cosmological models in CECG.

The Lagrangian we are going to analyze is
\begin{equation} \label{eq:ECGaction}
    S[g_{\mu\nu}] = \int \dd^4x \sqrt{|g|} \left(- \Lambda + \frac{\mpl^2}{2} \mathring{R} + \frac{\beta}{\mpl^2} \mathcal{R}_{(3)} \right)\,,
\end{equation}
where the General Relativity part has been corrected with the CECG invariant:\footnote{We keep the little ring over the curvatures to emphasize that we are in purely metric gravity, i.e., that all the curvatures are evaluated in the Levi-Civita connection.}
\begin{align}
    \mathcal{R}_{(3)} & = - \frac{1}{8} \Big(
    12 \mathring{R}_\mu{}^\rho{}_\nu{}^\sigma \mathring{R}_\rho{}^\tau{}_\sigma{}^\eta \mathring{R}_\tau{}^\mu{}_\eta{}^\nu  
    +\mathring{R}_{\mu\nu}{}^{\rho\sigma} \mathring{R}_{\rho\sigma}{}^{\tau\eta} \mathring{R}_{\tau\eta}{}^{\mu\nu}
    +2 \mathring{R} \mathring{R}_{\mu\nu\rho\sigma}\mathring{R}^{\mu\nu\rho\sigma}\nonumber \\
    & \qquad \qquad
    -8 \mathring{R}^{\mu\nu}\mathring{R}_\mu{}^{\rho\sigma \tau}\mathring{R}_{\nu\rho\sigma\tau}
    +4 \mathring{R}^{\mu\nu}\mathring{R}^{\rho\sigma}\mathring{R}_{\mu\rho\nu\sigma}
    -4 \mathring{R} \mathring{R}_{\mu\nu}\mathring{R}^{\mu\nu}
    +8 \mathring{R}_\mu{}^\nu \mathring{R}_\nu{}^\rho \mathring{R}_\rho{}^\mu
    \Big)\,.
\end{align}
Let us insist one more time that this combination has the same linear spectrum as GR around cosmological backgrounds (i.e., only the graviton propagates). 

For a FLRW universe of the type
\begin{equation}
  \dd s^2=\dd t^2 -a^2(t) \delta_{ij} \dd x^i\dd x^j\,, \label{eq:FLRW}
\end{equation}
the gravitational Friedmann equation is
\begin{equation}\label{eq: Isotropic condition}
3\mpl^2\mathrm{H}^2(t)-6\frac{\beta}{\mpl^2} \mathrm{H}^6(t)=\rho+\Lambda\,,
\end{equation}
where $\rho$ is the energy-density of the matter sector and $\mathrm{H}(t)\coloneqq \frac{\dot a}{a}$. In the absence of matter, $\rho=0$, we can see that we have (at most) three branches of expanding de Sitter solutions. For metric perturbations $g_{ij}=a^2(\delta_{ij}+h_{ij})$ with $h_{ij}$ transverse and traceless, we find the quadratic action 
\cite{PookkillathDeFelice2020}:
\begin{equation} \label{Eq:actiontensormodes}
S^{(2)}= \frac{\mpl^4-6 \mathrm{H}^4_0\beta}{8 \mpl^2} \sum_{\lambda} \int \dd t \, \dd^3x \, a^3 \left[ \dot{h}^2_\lambda- \frac{1}{a^2} (\partial_i h_\lambda)^2\right],
\end{equation}
where the sum extends to the two polarizations of the gravitational waves and $\mathrm{H}_0$ is the considered de Sitter branch. It is clear that $\mpl^4 - 6 \mathrm{H}^4_0\beta>0$ is required to avoid ghostly gravitational waves and, consequently, the only allowed de Sitter solutions verify $\Lambda > 2 \mathrm{H}^2_0>0$ \cite{PookkillathDeFelice2020}.

\subsubsection{Generalities about Bianchi I spaces}

To prove that spatially flat FLRW solutions are singular in CECG, we make use of a general Bianchi I metric. These spaces are anisotropic extensions of the flat FLRW spaces (the {\it isotropic limit}):
\begin{equation}
\dd s^2=\bar{g}_{\mu\nu}\dd x^\mu \dd x^\nu=\mathcal{N}^{\,2}(t)\dd t^2-a^2(t) \dd x^2-b^2(t) \dd y^2-c^2(t) \dd z^2  \,, \label{eq: BianchiI}
\end{equation}
where $\mathcal{N}(t)$ is the lapse function and $a(t)$, $b(t)$ and $c(t)$ stand for the scale factors along the three coordinate axis. The idea would be to study the dynamics very close to the isotropic solution.

First let us derive a couple of results valid for any gravitational action $S[g_{\mu\nu}]$. It is convenient to define the isotropic expansion rate
\begin{equation}\label{eq:defH}
\mathrm{H}(t)\coloneqq \frac{1}{3}\bigg(\frac{\dot{a}}{a}+\frac{\dot{b}}{b}+\frac{\dot{c}}{c}\bigg)\,.
\end{equation}
The deviation with respect to the isotropic case will be encoded in two functions, $\sigma_1(t)$ and $\sigma_2(t)$, defined implicitly by 
\begin{equation} \label{eq: def sigmas}
\frac{\dot{a}}{a}= \mathrm{H} + \epsilon_{\sigma} (2\sigma_1-\sigma_2)\,, \qquad 
\frac{\dot{b}}{b}= \mathrm{H} - \epsilon_{\sigma} (\sigma_1-2\sigma_2)\,, \qquad 
\frac{\dot{c}}{c}= \mathrm{H} - \epsilon_{\sigma} (\sigma_1+\sigma_2)\,,
\end{equation}
where $\epsilon_{\sigma}$ is a certain (not necessarily small) fixed parameter controlling the deviation ($\epsilon_{\sigma}=0$ for the isotropic case). 

The metric \eqref{eq: BianchiI} fulfills the requirements of the \emph{Palais' principle of symmetric criticality} \cite{Palais1979} (see also \cite{FelsTorre2002,DeserTekin2003,DeserFranklinTekin2004}), so one can use the minisuperspace approach
 and substitute the Ansatz \eqref{eq: BianchiI} in the action and then vary with respect to the free functions $\{\mathcal{N},a,b,c\}$,
\begin{equation}
  \bar{S} [\mathcal{N}, a, b, c] \coloneqq S[\bar{g}{}_{\mu\nu}(\mathcal{N}, a, b, c)]\,.
\end{equation}
It can be shown that, thanks to the Noether identity under diffeomorphisms, the full dynamics is determined by three equations, which can be chosen to be
\begin{equation} 
  0  =\frac{\delta \bar{S}}{\delta \mathcal{N}}\,,\qquad
  {\rm E}_1\coloneqq ~~ 0  = \frac{\delta \bar{S}}{\delta c}c - \frac{\delta \bar{S}}{\delta b}b\,,\qquad
  {\rm E}_2 \coloneqq ~~ 0  = \frac{\delta\bar{S}}{\delta c}c-\frac{\delta\bar{S}}{\delta a}a\,.
\end{equation}
After (and only after!) deriving the equations of motion in the minisuperspace approach, we can take cosmic time $\mathcal{N}=1$.

\subsubsection{Dynamical equations for the anisotropies: strong coupling issues}

In the theory \eqref{eq:ECGaction}, the highest-order derivatives of the anisotropy functions $\sigma_1$ and $\sigma_2$ appear in the evolution equations $\{{\rm E}_1,{\rm E}_2\}$, which indeed trivialize in the isotropic case. If we drop the global factor $\epsilon_{\sigma}$, they can be written in the following schematic form:
\renewcommand\arraystretch{1.3}
\begin{equation}\label{eq: principal part}
  \frac{\beta}{\mpl^2} \left[ \epsilon_{\sigma}
    {\bf M_1}
    \begin{pmatrix}\dddot{\sigma}_2 \\\dddot{\sigma}_1 \end{pmatrix}
    + \epsilon_{\sigma} 
    {\bf M_2} \begin{pmatrix}\ddot{\sigma}_2 \\\ddot{\sigma}_1 \end{pmatrix}+  {\bf V} \right] 
    +3 \mpl^2 \begin{pmatrix}3\mathrm{H}\sigma_2+\dot{\sigma}_2 \\ 3\mathrm{H}\sigma_1+\dot{\sigma}_1 \end{pmatrix} =0 \,.
\end{equation}\renewcommand\arraystretch{1}
where the matrices ${\bf M}_{1}$ and  ${\bf M}_{2}$, and the column vector ${\bf V}$ start at zeroth order in $\epsilon_{\sigma}$.\footnote{
    The components of ${\bf M}_1$ depend polynomially on $\sigma_1$, $\sigma_2$ and $\mathrm{H}$, whereas those of ${\bf M}_2$ and ${\bf V}$ also depend on $\dot{\sigma}_1$, $\dot{\sigma}_2$ and the derivatives of $\mathrm{H}$.} 
Notice that the higher-order terms containing second and third derivatives of the shear trivialize in the isotropic limit $\epsilon_{\sigma}\rightarrow0$. Consequently, in this limit, the order of the system of differential equations is abruptly reduced. In other words, we are losing d.o.f.'s (those associated to the higher-order derivative nature of the Lagrangian) in such a limit.

If we take for instance any Bianchi I metric, expand it around one of the de Sitter solutions (which is isotropic), it can be checked that order by order the resulting equations are of second order. So, around an isotropic background, the perturbative approach is not valid because it is not capturing the higher-order derivative nature of the full dynamics. Let us see this more clearly in an example:

\begin{boxmathclear}
\noindent{\bf Example.}

Consider a differential equation of the type (it should be understood as an schematic expression, since some factors must be introduced to have the appropriate dimensions):
\begin{equation}
    0= E := A(f) \dddot{f} + \ddot{f} + f^2\,.
\end{equation}
This equation is of third order in derivatives. Let us now perform an expansion  $f=f_0+f_1+f_2 + ...$ (we will ignore terms of order greater than 2), where $f_0$ is a known solution of the theory. The function $A$ can also be split according to the previous expansion, $A=A_0+A_1+A_2+...$ (notice that $A_0=A(f_0)$).

If we do the same with the equation $E=E_0+E_1+E_2+...$, we get
\begin{align}
    E_0 &= A_0 \dddot{f}_0 + \ddot{f}_0 + f^2_0\\
    E_1 &= A_0 \dddot{f}_1 +A_1 \dddot{f}_0 + \ddot{f}_1 + 2f_0f_1\\
    E_2 &= A_0 \dddot{f}_2 +A_1 \dddot{f}_1 + A_2 \dddot{f}_0 + \ddot{f}_2 + f_1^2+ 2f_0f_2\,.
\end{align}
Here, the equation $E_i$ should be seen as a differential equation for $f_i$, in which we have substituted $f_j$ with $j<i$ from the previous ones (i.e., we solve order by order). In particular, since $f_0$ is a solution of the theory the first equation is identically fulfilled. In principle, we see that all the equations continue being of third order for each of the $f_i$. But this is only true if $A_0\neq 0$, i.e., if the coefficient in front of the highest-order derivative term in the original equation does not vanish on the background we are considering. In fact, if $A_0=0$, the equations for the perturbations become of a lower order (the unknown in each case has been written in blue):
\begin{align}
    E_1 &= A_1 \dddot{f}_0 + {\color{blue}\ddot{f}_1} + 2f_0{\color{blue}f_1}\\
    E_2 &= A_1 \dddot{f}_1 + A_2 \dddot{f}_0 + {\color{blue}\ddot{f}_2} + f_1^2+ 2f_0{\color{blue}f_2}\,.
\end{align}
 
\end{boxmathclear}

What we have seen in this example is exactly what happens for CECG around isotropic solutions. This indicates that the perturbative expansion does not capture all of the information about the solution. In more mathematical terms, it can be seen that the isotropic solution belongs to a singular surface in phase space. In fact, in \cite{BeltranJCA2021} it was shown that it belongs to the intersection of the four singular surfaces of the equation.

The best way to verify that this situation leads to unstable behaviors is by looking at the the evolution of the Hubble parameter $\mathrm{H}(t)$ for randomly generated (and small) initial shears $\sigma_1$ and $\sigma_2$. As shown in Figure 2 and 3 of \cite{BeltranJCA2021}, the evolution curves of $\mathrm{H}$ deviate with respect to the isotropic (unperturbed) cases, clearly indicating that the isotropic background receives important corrections from very small anisotropies.

\subsection{Ghosts in Ricci-based gravity theories}
\label{sec:RB}

In this section we will argue why metric-affine theories that modify GR by adding higher-order curvature corrections are generically plagued by ghosts. More concretely, we will prove that theories where the action is a general function of the inverse metric and the Ricci tensor propagate ghost-like degrees of freedom unless projective symmetry is imposed, in which case they are GR in disguise \cite{BeltranJimenez:2017doy,Delhom:2021bvq}. Consider a metric-affine theory described by the action
\begin{equation}
    S[g^{\mu\nu},\Gamma^\alpha{}_{\mu\nu},\psi]=\frac{\mpl^2}{2}\int\dd^Dx \sqrt{-g}F(g^{\mu\nu},R_{\mu\nu})+\int\dd^Dx \sqrt{-g}\mathcal{L}_{\textrm{m}}(g^{\mu\nu},R_{\mu\nu},\psi)
    \label{eq:RBGAction}
\end{equation}
where the Ricci tensor is that of a generic affine connection and $\psi$ stands for a collection of matter fields, which may couple to the connection through the Ricci tensor.\footnote{This can be easily generalized to arbitrary couplings with the connection by just introducing a hypermomentum piece in the corresponding equations, see \cite{Delhom:2021bvq}.} This family of theories is known as \emph{(generalized) Ricci-based gravity} theories (gRBGs).

The dynamics for a theory of this family is obtained by varying the action with respect to metric, affine connection and matter fields independently. Particularly, the connection field equations are obtained by setting
\begin{equation}
    \int\dd^Dx \sqrt{-g}\left(\frac{\mpl^2}{2}\frac{\partial F}{\partial R_{\mu\nu}}+\frac{\partial\mathcal{L}_{\textrm{m}}}{\partial R_{\mu\nu}}\right)\delta_\Gamma R_{\mu\nu}\equiv\frac{\mpl^2}{2}\int\dd^Dx \sqrt{-q}q^{\mu\nu}\delta_\Gamma R_{\mu\nu}=0.
\end{equation}
This requirement leads to the connection equations
\begin{equation}
\nabla_{\lambda}\left[\sqrt{-q} q^{\nu \mu}\right]-\delta^{\mu}{}_{\lambda} \nabla_{\rho}\left[\sqrt{-q} q^{\nu \rho}\right]=\sqrt{-q}\left[T^{\mu}{}_{\lambda \alpha}q^{\nu \alpha}+T^{\alpha}{}_{\alpha \lambda} q^{\nu \mu}-\delta^{\mu}{}_{\lambda} T^{\alpha}{}_{\alpha \beta} q^{\nu \beta}\right]
\label{eq:ConnEqRBG}
\end{equation}
which are formally identical to the ones obtained for GR if we substitute the metric for a second rank tensor $q^{\mu\nu}$ with no symmetries. In the case of GR, the solution to these connection equation is given by the Levi-Civita connection for the metric (plus a gauge choice of projective mode). In the case of adding matter that couples to the connection algebraically, an extra term on the right called hypermomentum will arise which will make the solution depart from the Levi-Civita connection of the metric. However, having no symmetry in the indices of $q^{\mu\nu}$, the above connection equations are formally identical to those of Non Symmetric Gravity \cite{Moffat:1978tr,Moffat:1994hv},  a theory that is known by propagating ghostly degrees of freedom \cite{Damour:1991ru,Damour:1992bt} (see also \cite{BeltranDelhom2019,BeltranDelhom2020}). In order to benefit from this results, let us introduce the Einstein-like frame of this theories through the use of field redefinitions and the properties of auxiliary fields.

\subsubsection{Einstein-like frame for gRBGs and projective symmetry}

Let us introduce an auxiliary field $\Sigma_{\alpha\beta}$ that allows to linearize the action with respect to the Ricci tensor as
\begin{equation}
\begin{split}
S[g,\Gamma,\Sigma,\psi]=\frac{\mpl^{2}}{2}\int \dd^Dx \sqrt{-g}\Bigg[&F(g^{\mu\nu},\Sigma_{\mu\nu})+\frac{2}{\mpl^2}\mathcal{L}_\textrm{m}(g^{\mu\nu},\Sigma_{\mu\nu},\psi)\\
&+\left(\frac{\partial F}{\partial \Sigma_{\mu\nu}}+\frac{2}{\mpl^2}\frac{\partial \mathcal{L}_\textrm{m}}{\partial \Sigma_{\mu\nu}}\right)\big(R_{\mu\nu}-\Sigma_{\mu\nu}\big)\Bigg],
\end{split}
\label{eq:AuxAction}
\end{equation}
where, in some places, indexes are not explicitly written to lighten the notation. The equivalence between the above action and the original one can be seen by computing the field equations for $\Sigma$, which yield
\begin{equation}
    \lrsq{\frac{\partial^2 }{\partial \Sigma_{\mu\nu}\Sigma_{\alpha\beta}}\lr{F+\frac{2}{\mpl^2}\mathcal{L}_\textrm{m}}}\lr{R_{\mu\nu}-\Sigma_{\mu\nu}}=0
\end{equation}
which are algebraic equation that are uniquely solved by $\Sigma_{\mu\nu}=R_{\mu\nu}$ provided that the second derivative of $F+2\mathcal{L}_\textrm{m}/\mpl^2$ with respect to $\Sigma$ does not vanish identically. Therefore, $\Sigma$ is indeed auxiliary and by plugging its value in the action \eqref{eq:AuxAction} we recover the original one, thus showing that the above is just a reformulation of the original theory with an extra auxiliary field, and therefore physically equivalent.

Now, we can perform a field redefinition in the new action by defining a new field variable 
\begin{equation}
    \sqrt{-q}q^{\mu\nu}=\sqrt{-g}\frac{\partial}{\partial\Sigma_{\mu\nu}}\lr{F+\frac{2}{\mpl^2}\mathcal{L}_\textrm{m}}
\end{equation}
and we can replace $\Sigma$ by $q$ by formally inverting this algebraic relation above which yields solutions of the schematic form $\Sigma(q,g,\psi)$. The result is that the action \eqref{eq:AuxAction} now reads
\begin{equation}
S[g,\Gamma,q,\psi]=\frac{\mpl^{2}}{2}\int \dd^Dx \sqrt{-q}\Big[q^{\mu\nu}R_{\mu\nu}+U(g,q,\psi)\Big]
\label{eq:AuxActionConn}
\end{equation}
with $U(g,q,\psi)$ given by
\begin{equation}
    U(g,q,\psi):=\left.\frac{\sqrt{-g}}{\sqrt{-q}}\Bigg[F(g,\Sigma)+\frac{2}{\mpl^2}\mathcal{L}_\textrm{m}(g,\Sigma,\psi)-\left(\frac{\partial F}{\partial \Sigma_{\mu\nu}}+\frac{2}{\mpl^2}\frac{\partial \mathcal{L}_\textrm{m}}{\partial \Sigma_{\mu\nu}}\right)\Sigma_{\mu\nu}\Bigg]\right|_{\Sigma(q,g,\psi)}
\end{equation}
From \eqref{eq:AuxActionConn}, we see that the connection equations will be exactly the same as \eqref{eq:ConnEqRBG}. At this point, we can also get rid of the metric $g$ because it is also auxiliary, as can be seen by computing its field equations
\begin{equation}
    \frac{\partial{U}}{\partial g^{\mu\nu}}=0,
\end{equation}
which are clearly algebraic in $g$. Hence, we can solve them to find an algebraic solution for $g(q,\psi)$ in terms of the new object $q$ and the matter fields, and plugging it back into the action we arrive to
\begin{equation}
S[\Gamma,q,\psi]=\frac{\mpl^{2}}{2}\int \dd^D x \sqrt{-q}q^{\mu\nu}R_{\mu\nu}+\int \dd^D x\sqrt{-q}\mathcal{L}^\textrm{EF}_\textrm{m}(q,\psi),\label{eq:RBGActionEF}
\end{equation}
where we have defined the Einstein-frame matter action
\begin{equation}
    \mathcal{L}^\textrm{EF}_\textrm{m}(q,\psi):= \left.\frac{\mpl^2}{2}U(g,q,\psi)\right|_{g(q,\psi)}.
\end{equation}
We thus see that, with the machinery of field redefinitions and auxiliary fields, we can write any theory of the original gRBG form \eqref{eq:RBGAction} equivalently as \eqref{eq:RBGActionEF}, where the gravitational sector is reduced to a first-order Einstein-Hilbert-like term with the object $q$ acting as the metric, and a matter action where the same matter fields that were in the beginning now interact nonlinearly among themselves and couple also to $q^{\mu\nu}$.

However, the analogy is not completely satisfied in general because the Ricci tensor of a general affine connection has an antisymmetric piece, and therefore, so does the auxiliary field $\Sigma$ and by extension the object $q^{\mu\nu}$. Hence, in the general case, the Einstein frame of the theory is equivalent to the Non Symmetric Gravity theory formulated by Moffat as explained above. This theory is known by propagating extra ghostly degrees of freedom beyond the usual graviton around arbitrary backgrounds, a feature that spoils its physical viability as a description of our universe.

In the case of gRBG theories, there is a safe way of getting rid of this degrees of freedom that is implemented by requiring a symmetry in the affine sector so that the pathological degrees of freedom will not be reintroduced by quantum corrections (unless there is a gauge anomaly). This symmetry that can ghost-bust gRBG theories is called projective symmetry, and it is realized by having invariance under projective transformations,\footnote{
     The name of this transformations comes from the fact that they do not alter the family of autoparallel curves of the connection (they only introduce a reparameterization of them), hence preserving the affine structure locally introduced in each tangent space by the connection.}
which in local coordinates are given by
\begin{equation}
\Gamma^\alpha{}_{\mu\nu}\overset{\xi}{\longmapsto} \Gamma^\alpha{}_{\mu\nu}+\xi_\mu \delta^\alpha{}_\nu,
\label{eq.ProjTransf}
\end{equation}
where $\xi$ is an arbitrary vector field called projective mode. This transformation of the connection leads to a transformation in the Ricci tensor as
\begin{equation}
R_{\mu\nu}\overset{\xi}{\longmapsto} R_{\mu\nu}-2\partial_{[\mu}\xi_{\nu]},
\end{equation}
so that its symmetric part is invariant but its antisymmetric part changes with the fieldstrength of the projective mode. Hence, we see that imposing projective symmetry in gRBG theories amounts to require that only the symmetric piece of the Ricci tensor appears in the action. If we do so, then the auxiliary field $\Sigma$, and therefore the object $q$, are now both symmetric, and the analogy between the Einstein frame action \eqref{eq:RBGActionEF} and first-order GR is exactly fulfilled, where $q^{\mu\nu}$ is now a common metric and the connection is given by its Levi-Civita connection plus a spurious projective gauge mode as dictated by the dynamics.

On the other hand, lifting the requirement of projective symmetry allows $q$ to have an antisymmetric part, so that it can be decomposed as
\begin{equation}
\sqrt{-q}q^{\mu\nu}=\sqrt{-h}(h^{\mu\nu}+B^{\mu\nu}),
\end{equation} 
where $h^{\mu\nu}$ is symmetric and acts as a usual metric, and $B^{\mu\nu}$ is a 2-form that describes the antisymmetric part of $q^{\mu\nu}$. Thus, beyond the projective mode that will now acquire dynamics due to an explicit breaking of projective symmetry in the action, the 2-form field describing the antisymmetric part of the object $q$ will also introduce new degrees of freedom in general. Furthermore, note that the connection equations do not have the Levi-Civita connection of $\bar{q}$ as a solution anymore, as they will also depend on derivatives of the 2-form $B$. This will end up generating pathological couplings between the 2-form and the curvature of $h$ which lead to the propagation of Ostrogradski instabilities. In the following sections, we will see how these extra degrees of freedom are indeed ghostly around arbitrary backgrounds, spoiling the viability of gRBG theories without projective symmetry.

\subsubsection{Additional degrees of freedom: Ghosts from splitting the connection}
One way of showing the pathological nature of the new degrees of freedom that arise if projective symmetry is explicitly broken by allowing the antisymmetric piece of the Ricci tensor into the action is by splitting the connection into different pieces. First, note that for any affine connection $\Gamma^\alpha{}_{\mu\nu}$ and symmetric invertible 2nd rank tensor $h^{\mu\nu}$, we can always make the splitting
\begin{equation}
    \Gamma^\alpha{}_{\mu\nu}={}^h\Gamma^\alpha{}_{\mu\nu}+\Upsilon^\alpha{}_{\mu\nu},
\end{equation}
where ${}^h\Gamma$ is the Levi-Civita connection of $h^{\mu\nu}$ and $\Upsilon$ is a rank 3 tensor that encodes the difference between the full affine connection and ${}^h\Gamma$.\footnote{
    Note that, though the connection symbols are not tensors, the difference of two connections is always a tensor field.} 
Then, we can also strip a projective mode from $\Upsilon$ in order to explicitly see how it behaves when projective symmetry is explicitly broken. A convenient choice of stripping is
\begin{equation}
\upsilonh^\alpha{}_{\mu\nu}=\Upsilon^\alpha{}_{\mu\nu}+\frac{1}{D-1}\Upsilon_\mu\delta^\alpha{}_\nu,
\end{equation}
where $\Upsilon_\mu:=2\Upsilon^\alpha{}_{[\alpha\mu]}$. The convenience of this choice stems from the fact that the stripped connection $\upsilonh$ satisfies  $\upsilonh^{\alpha}{}_{[\alpha\mu]}=0$ and would allow to solve the connection equations if the 2-form $B$ vanished. Introducing this splitting explicitly in the action \eqref{eq:RBGActionEF} we find
\begin{equation}\label{eq:RBGNoProjectiveEinsteinFrameDecomposed}
\begin{split}
S[&h,B,\upsilonh,\Upsilon,\psi]=\frac{\mpl^2}{2}\int\dd^Dx\sqrt{-h}\Big[R^h-\frac{2}{D-1}B^{\mu\nu}\partial_{[\mu}\Upsilon_{\nu]}-B^{\mu\nu}\nabla^h_\alpha\upsilonh^\alpha{}_{\mu\nu}-B^{\mu\nu}\nabla^h_\nu\upsilonh^\alpha{}_{\alpha\mu}\\
&+\upsilonh^\alpha{}_{\alpha\lambda}\upsilonh^\lambda{}_{\kappa}{}^\kappa-\upsilonh^{\alpha\mu\lambda}\upsilonh_{\lambda\alpha\mu}-\upsilonh^\alpha{}_{\alpha\lambda}\upsilonh^\lambda{}_{\mu\nu}B^{\mu\nu}-\upsilonh^\alpha{}_{\nu\lambda}\upsilonh^\lambda{}_{\alpha\mu}B^{\mu\nu}\Big]+\tilde{S}^\textrm{EF}_\textrm{m}[h,B,\psi],
\end{split}
\end{equation}
where $R^h$ is the curvature scalar of $h_{\mu\nu}$ and $\tilde{S}^\textrm{EF}_\textrm{m}$ is just the matter action after the splitting. From this action, we readily see that the only kinetic term for projective mode $\Upsilon_\mu$ occurs through a coupling with $B$, so that around arbitrary $B$ backgrounds, this will render the vector unstable. To make this more explicit, note that the projective mode is oblivious to $\upsilonh$, and let us first consider the relevant sector of the action to describe it around a trivial 2-form background
\begin{equation}
\begin{split}
S\supset\int\dd^Dx\sqrt{-h}\Big[B^{\mu\nu}\partial_{[\mu}\Upsilon_{\nu]}-m^2 B_{\mu\nu}B^{\mu\nu}\Big],
\end{split}
\end{equation}
where the mass term will generally be present in $\tilde{S}^\textrm{EF}_\textrm{m}$ due to the old $U(q,g,\psi)$ after integrating out $g$ and splitting $q$ into $h+B$. Note that some proportionality factor has been absorbed into $m^2$ (and the other couplings of the action if we think of the full one) without loss of generality. Now, we can diagonalize the kinetic sector by performing the linear field redefinition
\begin{equation}
B^{\mu\nu}\mapsto\hat{B}^{\mu\nu}+\frac{1}{2m^2}\partial^{[\mu}\Upsilon^{\nu]}\qquad\text{and}\qquad \Upsilon_\mu\mapsto 2m\Upsilon_\mu,
\end{equation}
which yields
\begin{equation}
\begin{split}
S\supset\int\dd^Dx\sqrt{-h}\Big[\partial^{[\mu}\Upsilon^{\nu]}\partial_{[\mu}\Upsilon_{\nu]}-m^2 \hat{B}_{\mu\nu}\hat{B}^{\mu\nu}\Big],
\end{split}
\end{equation}
making apparent that the projective mode is a ghostly vector due to the wrong sign of its kinetic term. One could wonder whether nontrivial backgrounds of $B$ could stabilize the projective mode, acting as a ghost condensate. To see that this cannot be the case, note that around an arbitrary background of $B$, the mass term that would be generated for $B$ will not be diagonal, but rather through a mass matrix $M^{\alpha\beta\mu\nu}$, such that the relevant piece of the action to describe the dynamics of the projective mode would now read
\begin{equation}
\begin{split}
S\supset\int\dd^Dx\sqrt{-h}\Big[B^{\mu\nu}\partial_{[\mu}\Upsilon_{\nu]}-m^2 M^{\alpha\beta\mu\nu}B_{\mu\nu}B^{\mu\nu}\Big]
\end{split}
\end{equation}
so that the diagonalization of the kinetic sector is carried by the field redefinition
\begin{equation}
B^{\mu\nu}\mapsto\hat{B}^{\mu\nu}+\frac{1}{2m^2}\Lambda^{\mu\nu\alpha\beta}\partial_{[\alpha}\Upsilon_{\beta]}\qquad\text{and}\qquad \Upsilon_\mu\mapsto 2m\Upsilon_\mu,
\end{equation}
with $\Lambda^{\mu\nu\alpha\beta}$ satisfying $M^{\alpha\beta\rho\sigma}\Lambda_{\rho\sigma}{}^{\mu\nu}=h^{\alpha[\mu}h^{\nu]\beta}$ for the diagonalization to be successful. After these linear field redefinitions, the relevant piece of the action reads
\begin{equation}
\begin{split}
S\supset\int\dd^Dx\sqrt{-h}\Big[\Lambda^{\alpha\beta\mu\nu}\partial_{[\alpha}\Upsilon_{\beta]}\partial_{[\mu}\Upsilon_{\nu]}-m^2 M^{\alpha\beta\mu\nu}\hat{B}_{\alpha\beta}\hat{B}_{\mu\nu}\Big],
\end{split}
\end{equation}
so that now the possible ghostly character of the vector field is encoded in the eigenvalues of $\Lambda^{\alpha\beta\mu\nu}$. Concretely, the ghosts will be avoided if the eigenvalues of $\Lambda$ are all negative. For this to be satisfied, $\Lambda$ needs to have the same signature as $-h^{\alpha[\mu}h^{\nu]\beta}$.\footnote{
    Note that $h$ is a Lorentzian metric so that $h^{\alpha[\mu}h^{\nu]\beta}$ has positive eigenvalues.}
On the other hand, the stability of the 2-form mass term requires that the signature of $M^{\alpha\beta\mu\nu}$ is the same than that of $h^{\alpha[\mu}h^{\nu]\beta}$, namely that the mass matrix has positive eigenvalues. However, if the field redefinition diagonalizes the kinetic sector as above, we know that $M^{\alpha\beta\rho\sigma}\Lambda_{\rho\sigma}{}^{\mu\nu}=h^{\alpha[\mu}h^{\nu]\beta}$, so that both conditions cannot be satisfied at the same time, and the instability will appear either as a ghost in the vector sector or as a tachyon in the 2-form sector. Note, as well, that the field redefinition that diagonalizes the kinetic vector sector looks like a 2-form gauge transformation, so that the gauge invariant kinetic term for the 2-form is not affected and therefore its dynamics have no effect on the propagation of ghosts through the projective mode.

\subsubsection{Decoupling limits and the St\"uckelberg trick}

In the following section, we will unveil explicitly the presence of ghostly degrees of freedom in the decoupling limit for the 2-form field. Before that, let us clarify what we mean by decoupling limit in this short section. A \emph{decoupling limit} for a set of degrees of freedom, as the name indicates, is a limit taken in the parameters of the theory (masses, scales or couplings) in which these degrees of freedom decouple from the rest, so that some properties of the full theory may be easier to unveil. Note, however, that a decoupling limit represents only a certain regime of the theory, so some properties of the decoupling limit will not generalize to the full theory. For instance, being ghost-free in a decoupling limit does not mean that the theory is fully ghost-free, although finding ghosts in the decoupling limit typically implies that the ghosts will be in the full theory as well.

Decoupling limits are trivially implemented in some cases, think for example of Maxwell electrodynamics with charged fermions: by taking the limit of zero fine-structure constant, one arrives to a theory of a bunch of free fermions and a free massless vector field where all the degrees of freedom are decoupled. However, in some other cases, taking this limit in a physically sensible manner is not so trivial, as it may alter the number of degrees of freedom of the theory. This happens, for example, if one considers the massless limit of Proca theory, it propagates the two degrees of freedom that appear in a massles spin 1 theory, as opposed to the three degrees of freedom propagating in the full theory. In these cases one has to be more careful in taking the decoupling limit of the theory.

These problems appear when some gauge symmetries are recovered in such a limit, since they prevent some degrees of freedom from propagating. In these cases, a way of taking the decoupling limit is by introducing St\"uckelberg fields to restore the gauge symmetry in the full theory and then take their decoupling limit, which leads to the decoupling of the St\"uckelberg modes. This is sometimes referred to as \emph{St\"uckelberg trick}, and the result is a theory with a gauge field decoupled from the St\"uckelberg modes, which encode the extra degrees of freedom that are present of the full theory. Let us write down an explicit example on how to implement the St\"uckelberg trick for a Proca theory. 

\begin{boxmathclear}
\noindent \textbf{Example.}

Start from the Lagrangian
\begin{equation}\label{eq:Proca}
\mathcal{L}_{\textrm{Proca}}=-\frac14 F_{\mu\nu}F^{\mu\nu}+\frac12m^2A_\mu A^\mu.
\end{equation}
The mass term breaks the $U(1)$ gauge symmetry of the kinetic term since the transformation $A_\mu\mapsto \hat{A}_\mu+\partial_\mu\varphi$ leads to the addition of the terms $m^2 \hat{A}^\mu\partial_\mu\varphi$ and $(m^2/2)\partial_\mu\varphi\partial^\mu\varphi$ in the Lagrangian, which cannot be written as a total derivative. The St\"uckelberg trick here consists on two steps. 1) first restore the $U(1)$ gauge-invariance of the vector field by introducing the St\"uckelberg field $\varphi$ by making the replacement $A_\mu\mapsto \hat{A}_\mu+m^{-1}\partial_\mu\varphi$, which leads to the St\"uckelberg field Lagrangian \eqref{eq:ProcaActionStueckelberg} after a linear field redefinition $\varphi\mapsto m^{-1}\varphi$, namely
\begin{equation}
\mathcal{L}_\text{St\"uck}=-\frac{1}{4}\hat{F}_{\mu\nu}\hat{F}^{\mu\nu}+\frac{1}{2}m^{2}\hat{A}_{\mu}\hat{A}^{\mu}+m^2\hat{A}^\mu\partial_\mu\varphi+\frac{1}{2}\partial^\mu\varphi\partial_{\mu}\varphi\,.
\end{equation}
Here we see that the St\"uckelberg field is coupled to the massive vector through the vector's mass. In step 2) we consider the massless limit for the vector field, which is the decoupling limit for the St\"uckelberg mode, and yields
\begin{equation}
\mathcal{L}_\text{St\"uck}=-\frac{1}{4}\hat{F}_{\mu\nu}\hat{F}^{\mu\nu}+\frac{1}{2}\partial^\mu\varphi\partial_{\mu}\varphi\,,
\end{equation}
which indeed describes three degrees of freedom as the original theory, two encoded in the massless vector and one encoded in the St\"uckelberg field, which could be heuristically associated to the longitudinal polarization that is lost in the massless limit of the Proca theory. 

\end{boxmathclear}

This trick can be generalized in a straightforward manner to more general gauge theories, and we will use it in the next section to show how the the 2-form field $B$ and the projective mode propagate unstable degrees of freedom.
 
\subsubsection{Ghosts in the decoupling limit for the 2-form}

First, consider the 2-form sector perturbatively, so that from \eqref{eq:RBGActionEF} we find at quadratic order\footnote{In 4 dimensions and after a redefinition $B\mapsto 2B/\mpl^2$.}
\begin{align}
S^{(2)}=\int\dd^Dx\sqrt{-h}\Big[&\frac{\mpl^2}{2}R^{h}-\frac{1}{12} H_{\mu\nu\rho} H^{\mu\nu\rho}-\frac{1}{4} m^2 B^2-\frac{\sqrt{2}{\mpl}}{3}B^{\mu\nu}\partial_{[\mu}\Upsilon_{\nu]}\nonumber\\
&+\frac{1}{4} R^{h} B^2-R^{h}_{\mu\nu\alpha\beta}B^{\mu\alpha}B^{\nu\beta}\Big]
\label{eq:ActionPertB}
\end{align}
where $H_{\mu\alpha\beta}=\partial_{[\mu}B_{\alpha\beta]}$ is the fieldstrength of the 2-form, which provides a ghost-free (and gauge invariant) kinetic term for a 2-form. The mass of the 2-form is generated by the Einstein frame matter Lagrangian in general. The St\"uckelberg trick for the 2-form is parallel to that for the vector field, namely, introduce St\"uckelberg fields $b_\nu$ for the 2-form by the replacement 
\begin{equation}
B_{\mu\nu}\mapsto\hat{B}_{\mu\nu}+\frac{2}{m} \partial_{[\mu}b_{\nu]}.
\end{equation}
After introducing the St\"uckelberg field, performing the field redefinition 
\begin{equation}
\Upsilon_\mu\mapsto\frac{3m}{\sqrt{2}\mpl}\hat{\Upsilon}_\mu,
\end{equation}
and taking the Minkowskian limit, we arrive to the action
\begin{align}
S^{(2)}=\int\dd^Dx&\Big(-\frac{1}{12} \hat{H}_{\mu\nu\rho} \hat{H}^{\mu\nu\rho}-\frac{1}{4} m^2 \hat{B}^2-\partial^{[\mu}b^{\nu]}\partial_{[\mu}b_{\nu]}-m\hat{B}^{\mu\nu}\partial_{[\mu}b_{\nu]}\nonumber\\
&-m \hat{B}^{\mu\nu}\partial_{[\mu}\hat{\Upsilon}_{\nu]}-2\partial^{[\mu}b^{\nu]}\partial_{[\mu}\hat{\Upsilon}_{\nu]}\Big).
\end{align}
We can now take the decoupling limit for the St\"uckelberg as $m\rightarrow0$, which yields
\begin{align}
S^{(2)}=\int\dd^Dx&\Big(-\frac{1}{12} \hat{H}_{\mu\nu\rho} \hat{H}^{\mu\nu\rho}-\partial^{[\mu}b^{\nu]}\partial_{[\mu}b_{\nu]}-2\partial^{[\mu}b^{\nu]}\partial_{[\mu}\hat{\Upsilon}_{\nu]}\Big),
\end{align}
which describes the same number of degrees of freedom than a massive 2-form plus a massless vector, as the original action. Now, we can write the kinetic terms for the vectors as
 \begin{equation}
\begin{pmatrix}
\partial^{[\mu}b^{\nu]} & \partial^{[\mu}\hat{\Upsilon}^{\nu]}
\end{pmatrix}
\begin{pmatrix}
-1 & -1 \\
-1 & 0 
\end{pmatrix}
\begin{pmatrix}
\partial^{[\mu}b^{\nu]} \\
\partial^{[\mu}\hat{\Upsilon}^{\nu]}.
\end{pmatrix}
\end{equation} 
Therefore, we see that the kinetic matrix for the vector sector has eigenvalues $(-1\pm\sqrt{5})/2$. Since one is negative, this signals the presence of a ghostly degree of freedom, either in the St\"uckelberg field or in the projective mode. This can also be explicitly seen by diagonalizing the vectorial sector through the linear field redefinition
\begin{equation}
b_\mu\mapsto A_\mu+\xi_\mu\qquad\text{and}\qquad \Gamma_\mu\mapsto -2\xi_\mu,
\end{equation}
after which the second order action once the decoupling limit has been taken reads
\begin{align}
S^{(2)}=\int\dd^Dx&\Big(-\frac{1}{12} \hat{H}_{\mu\nu\rho} \hat{H}^{\mu\nu\rho}-\partial^{[\mu}A^{\nu]}\partial_{[\mu}A_{\nu]}+\partial^{[\mu}\xi^{\nu]}\partial_{[\mu}\xi_{\nu]}\Big),
\end{align}
where it can be seen that $\xi$ is a ghost around Minkowskian backgrounds. This will prevail for arbitrary backgrounds of $h^{\mu\nu}$ where, apart from the ghostly projective mode, we will also have problems in the 2-form sector due to its couplings to the curvature of the symmetric metric (see eq. \eqref{eq:ActionPertB}), which excite Ostrogradski instabilities through the appearance of non-degenerate 2nd order derivatives of $h$ in the action which will manifest as further ghosts in the 2-form sector. There are some technical details in this analysis that have been overlooked, such as the possibility of doing a quadratic field redefinition of the symmetric and antisymmetric piece of the metric. As well, interestingly, there are ways of getting rid of these pathologies within gRBG theories that relay on placing geometric constraints, rather than symmetry requirements. For a more detailed account of the instabilities that arise in gRBG without projective symmetry, as well as possible evasion mechanisms, or the impact of the analysis for more general metric-affine theories, the reader is referred to \cite{BeltranDelhom2020,Delhom:2021bvq}.

~

\begin{acknowledgments}
The authors are deeply grateful to Jose Beltr\'an Jim\'enez for his mentoring on the topic, the discussions about it, and feedback on the manuscript. We also thank Gerardo Garc\'ia-Moreno for useful comments. This work has been done under the support of the Estonian
Research Council and the European Regional Development Fund through the Center of Excellence TK133 ``The Dark Side of the Universe''. AJC was also supported by the Mobilitas Pluss post-doctoral grant MOBJD1035. FJMT is supported by the ``Fundaci\'on Ram\'on Areces''.
\end{acknowledgments}

\bibliographystyle{JHEP}
\bibliography{references.bib}

\end{document}